\documentclass[12pt]{iopart}
\pdfoutput=1
\usepackage{iopams}

\expandafter\let\csname equation*\endcsname\relax
\expandafter\let\csname endequation*\endcsname\relax
\usepackage{amsmath,amsfonts,amssymb}

\usepackage{siunitx}
\usepackage{tikz}
\usetikzlibrary{calc}
\usepackage{pgfplots}
\pgfplotsset{compat=1.11}

\usepackage{csquotes}
\usepackage{subcaption}
\usepackage{pgfplotstable}
\usepackage[noabbrev]{cleveref}
\usepackage{url}

\usepgfplotslibrary{external} 
\usepackage{natbib}
\setcitestyle{authoryear,open={(},close={)}}
\let\cite\citep

\usepackage{lineno}

\newcommand{\anon}[2]{#1} 

\begin{document}

\title[Exploration of Differentiability in a Proton CT Framework]{Exploration of Differentiability in a Proton Computed Tomography Simulation Framework}

\author{\anon{%
  Max Aehle\textsuperscript{a}, 
  Johan Alme\textsuperscript{b}, 
  Gergely Gábor Barnaföldi\textsuperscript{c}, 
  Johannes Blühdorn\textsuperscript{a}, 
  Tea Bodova\textsuperscript{b}, 
  Vyacheslav Borshchov\textsuperscript{d}, 
  Anthony van den Brink\textsuperscript{e}, 
  Viljar Eikeland\textsuperscript{b}, 
  Gregory Feofilov\textsuperscript{f}, 
  Christoph Garth\textsuperscript{g}, 
  Nicolas R.\ Gauger\textsuperscript{a}, 
  Ola Grøttvik\textsuperscript{b}, 
  Håvard Helstrup\textsuperscript{h}, 
  Sergey Igolkin\textsuperscript{f}, 
  Ralf Keidel\textsuperscript{i, a}, 
  Chinorat Kobdaj\textsuperscript{j}, 
  Tobias Kortus\textsuperscript{i}, 
  Lisa Kusch\textsuperscript{a}, 
  Viktor Leonhardt\textsuperscript{g}, 
  Shruti Mehendale\textsuperscript{b}, 
  Raju Ningappa Mulawade\textsuperscript{i}, 
  Odd Harald Odland\textsuperscript{k, b}, 
  George O'Neill\textsuperscript{b}, 
  Gábor Papp\textsuperscript{l}, 
  Thomas Peitzmann\textsuperscript{e}, 
  Helge Egil Seime Pettersen\textsuperscript{k}, 
  Pierluigi Piersimoni\textsuperscript{b, m}, 
  Rohit Pochampalli\textsuperscript{a}, 
  Maksym Protsenko\textsuperscript{d}, 
  Max Rauch\textsuperscript{b}, 
  Attiq Ur Rehman\textsuperscript{b}, 
  Matthias Richter\textsuperscript{n}, 
  Dieter Röhrich\textsuperscript{b}, 
  Max Sagebaum\textsuperscript{a}, 
  Joshua Santana\textsuperscript{i}, 
  Alexander Schilling\textsuperscript{i}, 
  Joao Seco\textsuperscript{o, p}, 
  Arnon Songmoolnak\textsuperscript{b, j}, 
  Ákos Sudár\textsuperscript{c}, 
  Ganesh Tambave\textsuperscript{q}, 
  Ihor Tymchuk\textsuperscript{d}, 
  Kjetil Ullaland\textsuperscript{b}, 
  Monika Varga-Kofarago\textsuperscript{c}, 
  Lennart Volz\textsuperscript{r, s}, 
  Boris Wagner\textsuperscript{b}, 
  Steffen Wendzel\textsuperscript{i}, 
  Alexander Wiebel\textsuperscript{i}, 
  RenZheng Xiao\textsuperscript{b, t}, 
  Shiming Yang\textsuperscript{b}, 
  Sebastian Zillien\textsuperscript{i}
}{(author names removed for double-anonymous peer review)} }

\address{%
  \anon{%
  \textsuperscript{a}Chair for Scientific Computing, University of Kaiserslautern-Landau, 67663 Kaiserslautern, Germany;
  \textsuperscript{b}Department of Physics and Technology, University of Bergen, 5007 Bergen, Norway;
  \textsuperscript{c}Wigner Research Centre for Physics, Budapest, Hungary;
  \textsuperscript{d}Research and Production Enterprise ``LTU'' (RPE LTU), Kharkiv, Ukraine;
  \textsuperscript{e}Institute for Subatomic Physics, Utrecht University/Nikhef, Utrecht, Netherlands;
  \textsuperscript{f}St.\ Petersburg University, St.\ Petersburg, Russia;
  \textsuperscript{g}Scientific Visualization Lab, University of Kaiserslautern-Landau, 67663 Kaiserslautern, Germany;
  \textsuperscript{h}Department of Computer Science, Electrical Engineering and Mathematical Sciences, Western Norway University of Applied Sciences, 5020 Bergen, Norway;
  \textsuperscript{i}Center for Technology and Transfer (ZTT), University of Applied Sciences Worms, Worms, Germany;
  \textsuperscript{j}Institute of Science, Suranaree University of Technology, Nakhon Ratchasima, Thailand;
  \textsuperscript{k}Department of Oncology and Medical Physics, Haukeland University Hospital, 5021 Bergen, Norway;
  \textsuperscript{l}Institute for Physics, Eötvös Loránd University, 1/A Pázmány P. Sétány, H-1117 Budapest, Hungary;
  \textsuperscript{m}FSN Department, ENEA, Frascati Research Center, 00044, Frascati, Italy;
  \textsuperscript{n}Department of Physics, University of Oslo, 0371 Oslo, Norway;
  \textsuperscript{o}Department of Biomedical Physics in Radiation Oncology, DKFZ—German Cancer Research Center, Heidelberg, Germany;
  \textsuperscript{p}Department of Physics and Astronomy, Heidelberg University, Heidelberg, Germany;
  \textsuperscript{q}Center for Medical and Radiation Physics (CMRP), National Institute of Science Education and Research (NISER), Bhubaneswar, India;
  \textsuperscript{r}Biophysics, GSI Helmholtz Center for Heavy Ion Research GmbH, Darmstadt, Germany;
  \textsuperscript{s}Department of Medical Physics and Biomedical Engineering, University College London, London, UK;
  \textsuperscript{t}College of Mechanical \& Power Engineering, China Three Gorges University, Yichang, People’s Republic of China
  }{(author address information removed for double-anonymous peer review)}
}
\ead{\anon{max.aehle@scicomp.uni-kl.de}{(e-mail address removed for double-anonymous peer review)}}
\vspace{10pt}
\begin{indented}
\item[]May 2023
\end{indented}

\begin{abstract}
\emph{Objective.} 
Algorithmic differentiation (AD) can be a useful technique to numerically optimize design and algorithmic parameters by, and quantify uncertainties in, computer simulations. However, the effectiveness of AD depends on how ``well-linearizable'' the software is. In this study, we assess how promising derivative information of a typical proton computed tomography (pCT) scan computer simulation is for the aforementioned applications. 
\emph{Approach.} This study is mainly based on numerical experiments, in which we repeatedly evaluate three representative computational steps with perturbed input values. We support our observations with a review of the algorithmic steps and arithmetic operations performed by the software, using debugging techniques.
\emph{Main results.}
 The model-based iterative reconstruction (MBIR) subprocedure (at the end of the software pipeline) and the Monte Carlo (MC) simulation (at the beginning) were piecewise differentiable. Jumps in the MBIR function arose from the discrete computation of the set of voxels intersected by a proton path. Jumps in the MC function likely arose from changes in the control flow that affect the amount of consumed random numbers. The tracking algorithm solves an inherently non-differentiable problem. 
\emph{Significance.} The MC and MBIR codes are ready for the integration of AD, and further research on surrogate models for the tracking subprocedure is necessary. 
\end{abstract}

%
\vspace{2pc}

\newcommand{\RR}{{\mathbb R}}
\newcommand{\hoch}[1]{^{\mathrm{(#1)}}}
\newcommand{\xo}{{\hat x}}
\newcommand{\io}{{\hat \imath}}
\newcommand{\jo}{{\hat \jmath}}
\tikzstyle{varblock}=[align=center,draw=black,rounded corners=5,font=\scriptsize]
\tikzstyle{funarrowA}=[rounded corners=10pt]
\tikzstyle{funarrowB}=[above,midway,font=\scriptsize,align=center]

\newcommand{\Cpp}{C\nolinebreak\hspace{-.05em}\raisebox{.4ex}{\tiny\bfseries +}\nolinebreak\hspace{-.10em}\raisebox{.4ex}{\tiny\bfseries +}}

\section{Introduction}\label{sec:introduction}

The option of treating cancer using beams of high-energy charged particles (mainly protons) is becoming more and more available on a world-wide scale. The main advantage over conventional x-ray radiotherapy lies in a possibly lower dose deposited outside the tumor, as the energy deposition of protons is concentrated around the so-called \emph{Bragg peak}. The depth of the Bragg peak depends on the beam energy and the \emph{relative stopping power} (RSP) of the tissue it traverses. Thus, treatment planning relies on a three-dimensional RSP image of the patient. In the state of the art calibration procedure of x-ray CT for proton therapy, scanner-specific look-up tables are used to convert the information retrieved from x-ray (single-energy or dual-energy) CT acquisitions into an RSP image: This approach comes with an uncertainty of the Bragg peak location of up to 3\% of the range \cite{Yang_2012,Paganetti_2012,wohlfahrt_2020}. On the other hand, the direct reconstruction of a \emph{proton CT} image using a high-energy proton beam and a particle detector has been shown to be intrinsically more accurate  \cite{dedes_experimental_2019,yang_theoretical_2010}.

To this end, the \anon{Bergen pCT collaboration \cite{alme_high-granularity_2020}}{XXX collaboration (name removed for double-anonymous peer review)} is designing and building a high-granularity digital tracking calorimeter (DTC) as a clinical prototype to be used as proton imaging device in existing treatment facilities for proton therapy. Its sensitive hardware consists of two \emph{tracking} and 41~\emph{calorimeter} layers of 108~ALPIDE (ALICE pixel detector) chips \cite{aglieri_rinella_alpide_2017} each. After traversing the patient, energetic protons will activate pixel clusters around their tracks in each layer until they are stopped, as shown in \cref{fig:schematic-scanning-process}. In each read-out cycle, the layer-wise binary activation images from hundreds of protons are collected and used to reconstruct the protons' paths and ranges through the detector and thus their residual direction and energy after leaving the patient. Based on this data from various beam positions and directions, a \emph{model-based iterative reconstruction} (MBIR) algorithm reconstructs the three-dimensional RSP image of the patient. The reconstructions of proton histories and of the RSP image are displayed as two main subprocedures in \cref{fig:pipeline-coarse}, along with a \emph{Monte Carlo simulation} subprocedure to generate the detector output instead of a real device for testing and optimization purposes.

\emph{Algorithmic differentiation} (AD) \cite{griewank_evaluating_2008,naumann_art_2011} is a set of techniques to efficiently obtain precise derivatives of a mathematical function given by a computer program. Such derivatives have been successfully used for optimization problems in various contexts, such as machine learning \cite{baydin_ad_in_ml_survey} and computational fluid dynamics \cite{Albring_etal2016b}, and AD is currently also adopted in the fundamental physics community for detector optimization \cite{baydin_toward_2021,dorigo_toward_2022}. Besides, particular methodologies for uncertainty quantification (UQ) involve derivatives. 
However, algorithmic derivatives can only be useful for optimization and UQ if the differentiated function, i.\,e.\ the pCT software pipeline, is sufficiently smooth. For the work presented in this article, we therefore study the reaction of three representative substeps of the pCT reconstruction process on changes in a single input parameter, keeping the other inputs fixed.

In \cref{sec:software-pipeline}, we summarize all the computational steps of the software pipeline in greater detail. \Cref{sec:ad} is a general introduction into the purpose and calculation of derivatives of computer programs. The numerical experiments are outlined in \cref{sec:methodology} and their results are stated in \cref{sec:results}. In \cref{sec:pipeline-again}, we analyze the observed discontinuous or non-differentiable behaviour and propose ways to mitigate it, and close with a summary and conclusions in \cref{sec:pipeline-again}.

\newcommand{\minisq}[2]{ \fill (#1-0.05, #2) -- +(0.1,0) -- +(0.1,0.1) -- +(0,0.1) -- cycle; }
\newcommand{\sqrow}[1]{
  \draw[very thin] (#1-0.05,-2) -- +(0,4);
  \draw[very thin] (#1+0.05,-2) -- +(0,4);
  \foreach \yl in {0,...,40}
    \draw[very thin] (#1-0.05,\yl*0.1-2) -- +(0.1,0);
}

\begin{figure}
    \centering
    \begin{tikzpicture}

\sqrow{0}
\sqrow{1}
\foreach \xl in {0,...,8}{
  \sqrow{2+\xl*0.2}
}
\minisq{0}{0.5}
\minisq{1}{0.2}
\minisq{1}{0.3}
\minisq{2.0}{0.1}
\minisq{2.2}{0.1}
\minisq{2.4}{0.1}
\minisq{2.6}{-0.1}
\minisq{2.6}{0}
\minisq{2.8}{-0.2}
\minisq{2.8}{-0.1}
\minisq{2.8}{0}
\minisq{2.8}{0.1}
\minisq{3.0}{0.1}
\minisq{3.0}{0.0}
\draw[blue,thick] (0,0.5) -- (1,0.3) -- (2,0.15) -- (2.2,0.13) -- (2.4,0.1) -- (2.6,0.04) -- (2.8,0.00) -- (3.0,0.1);

\draw (0,2) -- (0.5,3) node[anchor=south,align=center]{\scriptsize tracking \\[-0.2cm] \scriptsize layers} -- (1,2);
\draw (2.8,3) node[anchor=south,align=center] (calolabel) {\scriptsize calorimeter \\[-0.2cm] \scriptsize layers} ;
\foreach \xl in {0,...,8}
  \draw (2+\xl*0.2,2) -- (calolabel);

\draw[gray,fill=gray!30] (-2.8,0) ellipse(0.69*1.3 and 0.92*1.3);

\foreach \xl in {0,...,20} {
  \draw[very thin] (\xl*0.12-4.0,-1.2) -- (\xl*0.12-4.0,1.2);
}
\foreach \yl in {0,...,20} {
  \draw[very thin] (-4.0,\yl*0.12-1.2) -- (-1.6,\yl*0.12-1.2);
}

\fill[gray] (12*0.12-4.0, 2*0.12-1.2) -- +(0.12,0) -- +(0.12,0.12) -- +(0,0.12) -- cycle;
\draw ($(12*0.12-4.0, 2*0.12-1.2) + (0.06,0.0)$) -- +(0.5,-1) node[anchor=north,align=center]{\scriptsize reconstruct the \\[-0.2cm] \scriptsize RSP $\hat S(\vec x)$ at each voxel \\[-0.2cm] \scriptsize inside the object };

\draw (-5.8,0.4) node [above left,align=center] {\scriptsize beam delivery \\[-0.2cm] \scriptsize system};
\draw[thick, rotate around={10:(-5.8,0.4)}] (-6.1,0.3) -- (-5.5,0.3) -- (-5.5,0.5) -- (-6.1,0.5);

\begin{scope}[draw=blue,fill=blue,  thick]
\draw ($(-5.8,0.4)+({cos(10)*0.3},{sin(10)*0.3})$) -- +($({cos(10)*2.1},{sin(10)*2.1})$) coordinate (entrypoint) node[below=-0.1cm,midway,sloped,blue]{\scriptsize straight};
\draw (0,0.5) -- +($2.3*(-1,0.2)$) coordinate (exitpoint) node[below=-0.1cm,midway,sloped,blue]{\scriptsize straight};
\draw (entrypoint) .. controls ($(entrypoint)+({cos(10)*0.8},{sin(10)*0.8}) $)   and ($(exitpoint)+0.6*(-1,0.2)$)   .. (exitpoint) node[below=-0.05cm,midway,sloped,blue]{\scriptsize MLP};
\end{scope}
    
\end{tikzpicture}
    \caption{Schematic figure of the scanning process.}
    \label{fig:schematic-scanning-process}
\end{figure}
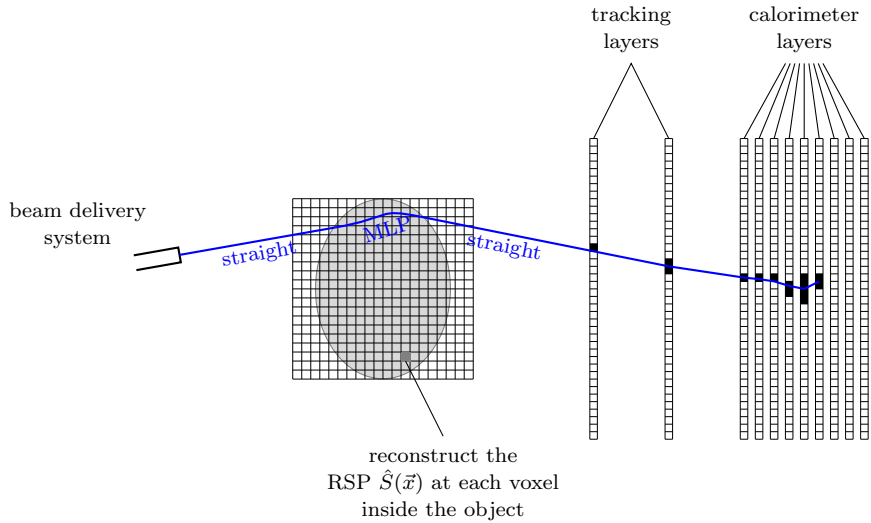

\section{Methods}
\subsection{Simulation of Proton CT Data Acquisition and Processing}\label{sec:software-pipeline}
\subsubsection{Foundations}

When energetic protons pass through matter, they slow down in a stochastic way, mainly due to inelastic interactions with the bound electrons. The average rate $\tfrac{\partial E}{\partial s}$ of kinetic energy $E$ lost per travelled length $s$ is called \emph{stopping power}. We denote it by $S(E,\vec x)$, indicating its dependency on the current energy $E$ of the proton and the local material present at location $\vec x$. Using the symbol $S_w(E)$ for the stopping power of water at the energy $E$, the RSP is defined as \begin{equation}\label{eq:def-rsp}
\hat S(\vec x) = \frac{ S(E, \vec x) }{ S_w(E) }.
\end{equation}
The dependency on $E$ has been dropped in the notation because in the relevant range between \SI{30}{\mega\eV} and \SI{200}{\mega\eV}, the RSP is essentially energy independent \cite{hurley_water-equivalent_2012}. 

Separating variables and integrating, one obtains
\begin{equation}\label{eq:ausr-wepl}
\int_{E(0)}^{E(\ell)} \frac{1}{S_w(E)} \,\mathrm d E = \int_{0}^{\ell} \hat S(\vec x(s)) \,\mathrm ds.
\end{equation}
This integral value is called the \emph{water-equivalent path length} (WEPL).

In list-mode, or single-event, pCT imaging, millions of protons are sent through the patient, and their positions, directions and energies are recorded separately, both before entering and after leaving the patient. In the setup conceived by the \anon{Bergen pCT}{XXX} collaboration and sketched in \cref{fig:schematic-scanning-process}, the exit measurements are performed by the \textit{tracking layers} of the DTC, through the processing steps outlined in \cref{sec:proton-recon}. While many prototypes reported in the literature use an additional pair of front trackers \cite{hurley_water-equivalent_2012,meyer_optimization_2020,esposito_pravda_2018,scaringella_prima_2013,saraya_study_2014,naimuddin_development_2016}, the setup at hand infers the positions and directions of entering protons from the beam delivery monitoring system. The employment of a single pair of tracking layers has been shown to be sufficiently accurate for dose-planning purposes through MC simulation studies \cite{solie_imagequality_2020}.

\begin{figure}
\centering
\begin{subfigure}{\textwidth}
\centering
\begin{tikzpicture}

\node (detectoroutput) at (0,0) [varblock] {detector \\ output };

\coordinate (geantinput) at ($(detectoroutput)-(2,0)$) ;
\node (originalrsp) at ($(geantinput)+(-1.3,0)$) [varblock] {orig.\ RSP};
\node (detectorparameters) at ($(geantinput)+(-0.5,1.4)$) [varblock] {detector parameters};
\node (randompath) at ($(geantinput)+(-0.7,-1)$) [varblock] {RNG};
\node (crosssections) at ($(geantinput)+(-1.6,0.7)$) [varblock] {physics param.};
\draw [funarrowA] (detectorparameters.south) -- (geantinput) -- (detectoroutput.west);
\draw [funarrowA] (originalrsp.east) -- (geantinput) -- (detectoroutput.west);
\draw [funarrowA] (randompath.north) -- (geantinput) -- (detectoroutput.west);
\draw [funarrowA] (crosssections.east) -- (geantinput) -- (detectoroutput.west);
\draw [->] (geantinput) -- (detectoroutput.west) node [funarrowB] {MC sim.\ };

\node (posanglewepl) at ($(detectoroutput)+(4.5,0)$) [varblock] {proton  positions, \\ directions,  WEPL};
\draw [->] (detectoroutput.east) -- (posanglewepl.west) node [funarrowB] {proton \\ reconstruction };

\node (reconstructedrsp) at ($(posanglewepl)+(4.5,0)$) [varblock] {recon.\ RSP};
\draw [->] (posanglewepl.east) -- (reconstructedrsp.west) node [funarrowB] {MBIR \\ subprocedure};

\end{tikzpicture}
\caption{Overview of the pCT reconstruction pipeline.}
\label{fig:pipeline-coarse}
\end{subfigure}
\\[1cm]
\begin{subfigure}{\textwidth}
\centering
\begin{tikzpicture}
\node (geantmac) at (0,0) [varblock] { GATE \\ .mac file };
\coordinate (geantinput) at ($(geantmac)-(2,0)$);
\node (originalrsp) at ($(geantinput)+(-1.3,0)$) [varblock] {orig.\ RSP};
\node (detectorparameters) at ($(geantinput)+(-0.5,1.4)$) [varblock] {detector parameters};
\node (crosssections) at ($(geantinput)+(-1.6,0.7)$) [varblock] {physics param.};
\draw [funarrowA] (detectorparameters.south) -- (geantinput) -- (geantmac.west);
\draw [funarrowA] (originalrsp.east) -- (geantinput) -- (geantmac.west);
\draw [funarrowA] (crosssections.east) -- (geantinput) -- (geantmac.west);
\draw[->] (geantinput) -- (geantmac.west) node [funarrowB] {.mac gen.\ };
  
\node (geantout) at ($(geantmac)+(3,0)$) [varblock] { GATE \\ .ROOT file };
\node (randompath) at ($(geantout)+(-2,-1)$) [varblock] {RNG};
\draw [->] (geantmac) -- (geantout) node [funarrowB] {GATE: \\ particle \\ transport };
\draw [funarrowA] (randompath.north) -- ($0.5*(geantmac.east)+0.5*(geantout.west)$) -- (geantout.west);

\node (detectoroutput) at ($(geantout)+(5.5,0)$) [varblock] { detector \\ output };
\draw [->] (geantout.east) -- (detectoroutput.west) node [funarrowB] { determine activated pixels \\ charge diffusion model } ;

\node (randomdiff) at ($(detectoroutput)+(-3,-1)$) [varblock] {RNG};
\draw [funarrowA] (randomdiff.north) -- ($0.5*(geantout.east)+0.5*(detectoroutput.west)$) -- (detectoroutput.west);

\draw [funarrowA] (detectorparameters.east) -- ($(geantout)+(1.0,1.4)$) -- ($(geantout)+(1.0,0)$) -- (detectoroutput.west);

\end{tikzpicture}
\caption{Computational steps of the Monte Carlo Simulation.}
\label{fig:monte-carlo}
\end{subfigure}
\\[1cm]
\begin{subfigure}{\textwidth}
\centering
\begin{tikzpicture}

\node (detectoroutput) at (0,0) [varblock] { detector \\ output };
\node (clusters) at ($(detectoroutput)+(3.5,0)$) [varblock] {clusters: \\ coordinates \\ and edeps} ;
\node (tracks) at ($(clusters)+(3.5,0)$) [varblock] {tracks: \\ grouped \\ clusters };
\node (wepls) at ($(tracks)+(4,-0.75)$) [varblock] {proton WEPLs $x\hoch{W}$};
\node [anchor=west] (posangle) at ($(wepls.west)+(0,1.5)$) [varblock] {proton positions, \\ directions $x\hoch{PD}$};

\draw [->] (detectoroutput.east) -- (clusters.west) node [funarrowB] {clustering};
\draw [->] (clusters.east) -- (tracks.west) node [funarrowB] {tracking};
\draw [->] (tracks.east) -- (wepls.west) node [funarrowB,sloped] {range calc.};
\draw [->] (tracks.east) -- (posangle.west) node [funarrowB,sloped] { ~ };
\end{tikzpicture}
\caption{Computational steps of the proton reconstruction subprocedure.}
\label{fig:cluster}
\end{subfigure}
\\[1cm]
\begin{subfigure}{\textwidth}
\centering
\begin{tikzpicture}
\node (posangle) at (0,0) [varblock] { proton positions,\\ directions $x\hoch{PD}$  };
\node [anchor=west] (wepl) at ($(posangle.west)+(0,-1.5)$) [varblock] {proton WEPLs $x\hoch{W}$ };
\node (matrixA) at ($(posangle)+(4,0)$) [varblock] {coefficient \\ matrix $A$ };
\node (reconstructed) at ($(matrixA)+(5,0)$) [varblock] {recon.\ RSP};

\draw[->] (posangle.east) -- (matrixA.west) node [funarrowB] {matrix\\generation};
\draw[->] (matrixA.east) -- (reconstructed.west) node [funarrowB] {solve \\ lin.\ system};
\draw[funarrowA] (wepl.east) -- ($(matrixA.east)+(0.3,-1.5)$) -- ($(matrixA.east)+(0.3,0)$) -- (reconstructed.west);

\end{tikzpicture}
\caption{Computational steps of the MBIR subprocedure.}
\label{fig:recon}
\end{subfigure}
\caption{Computational steps in the pCT reconstruction pipeline.}
\label{fig:computational-steps-big-figure}
\end{figure}
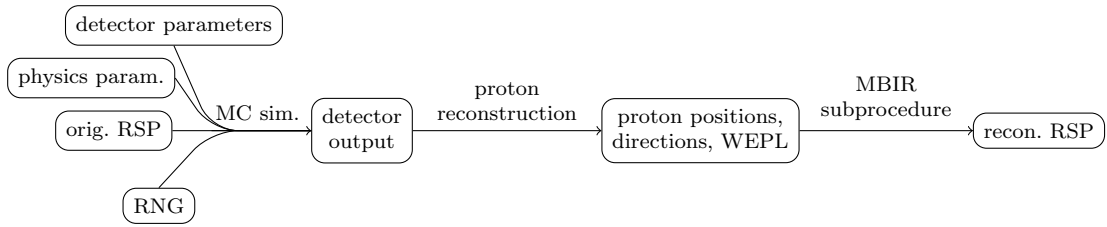
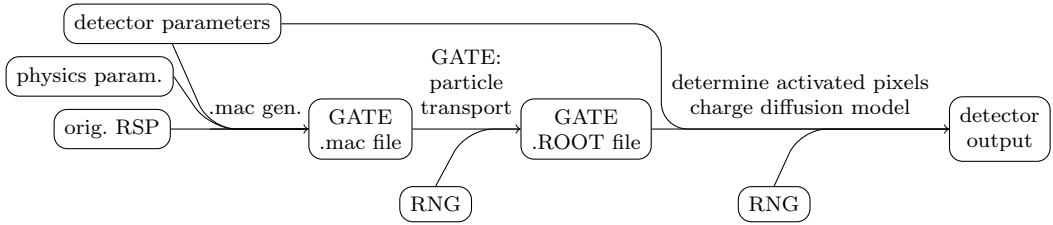
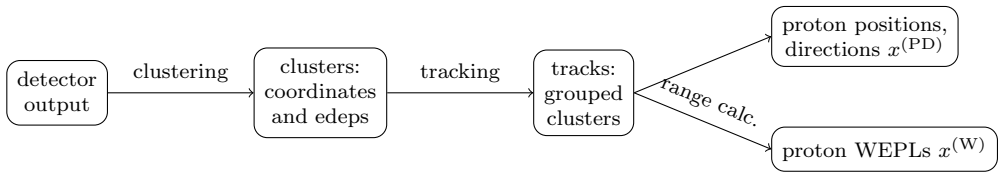
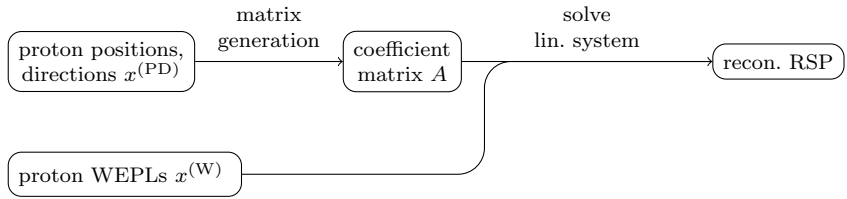

\subsubsection{Monte Carlo Subprocedure}\label{sec:mc}

\Cref{fig:monte-carlo} displays the intermediate variables and computational steps of the Monte Carlo subprocedure to simulate the detector output. Its central step is the open-source software GATE \cite{jan_gate_2004} for simulations in medical imaging and radiotherapy, based on the Geant4 toolkit for the simulation of the passage of particles through matter \cite{agostinelli_geant4simulation_2003,allison_geant4_2006,allison_recent_2016}. Based on a description of the relevant physical properties of the complete detector-patient setup, i.\,e., \begin{itemize}
\item the shape and material composition of the detector as well as the proton beam characteristics (\emph{detector parameters}),
\item the shape and material composition of the patient (\emph{original RSP}), and 
\item physics parameters like models and cross sections that define probability distributions for the relevant interactions between particles and matter, as well as output parameter setup (\emph{physics parameters}),
\end{itemize}
GATE produces stochastically independent paths of single particles through the described setup. Whenever interactions occur, with a certain probability distribution, the turnout for the proton at hand is decided by a random number from a pseudo-random number generator (RNG).

The epitaxial layers of the ALPIDE chips are modelled as single {\ttfamily crystalSD} volumes in GATE, so the positions and energy losses of all charged particles passing these volumes are recorded. %
In the real DTC, the energy deposition in each layer is estimated from the number of pixels activated by electron diffusion around the track, i.\,e.\ the magnitude of a pixel \textit{cluster} \cite{tambave_2020}. To reproduce this diffusion effect for the MC simulation, a random cluster, whose size corresponds to the recorded energy loss, is retrieved from a library containing the observed cluster shapes and their occurrence probability sampled from experimental data \cite{pettersen_design_2019}. 
As approximately 100 protons pass through the patient during one read-out cycle of the real detector (depending on the detector and beam parameters), a union of all activated pixels is formed to obtain the final binary image for each read-out cycle. 

\subsubsection{Proton Track Reconstruction}\label{sec:proton-recon}

\Cref{fig:cluster} displays the computational steps to convert the binary activation images per layer and read-out cycle, either produced by the real detector or by the Monte Carlo procedure in \cref{sec:mc}, back to continuous  coordinates and energies of protons.

As mentioned above, neighbouring (vertically and horizontally) activated pixels are grouped into \emph{clusters} per layer and read-out cycle. 
The proton's coordinate is given by the cluster's center of mass and its energy deposition is related to the size of the cluster \cite{pettersen_proton_2017, tambave_2020}. 

In the \emph{tracking} step, a track-following procedure \cite{strandlie_track_2010} attempts to match clusters in bordering layers likely belonging to the same particle trajectory. The angular change between the extrapolation of a growing track and cluster candidates in a given layer is minimized in a recursive fashion \cite{pettersen_design_2019,pettersen_proton_2020}.

Based on the cluster coordinates in the two tracking layers of the DTC, the position and direction of the proton exiting the patient can be inferred. A  vector $x\hoch{PD}$ stores this data together with the position and direction of the beam source.

The proton's residual WEPL before entering the detector can be estimated by a fit of the Bragg-Kleeman equation of Bortfeld \cite{bortfeld_analytical_1997,pettersen_accuracy_2018} to the energy depositions per layer \cite{pettersen_design_2019}. Its difference to the initial beam's WEPL is stored in $x\hoch{W}$. Failures of the tracking algorithms are usually due to pair-wise confusion between tracks from \emph{multiple Coulomb scattering} (MCS), to the merging of close clusters, or to high-angle scattering. To this end, tracks with an unexpected distribution of energy depositions are attributed to secondary particles or mismatches in the tracking algorithm, and filtered out \cite{pettersen_investigating_2021}.

\subsubsection{MBIR Subprocedure}\label{sec:mbir}
Model-based iterative reconstruction algorithms repeatedly update the RSP image to make it a better and better fit to the measurements of $x\hoch{PD}$, $x\hoch{W}$,  according to the model equations \eqref{eq:ausr-wepl}.

While the paths of the protons in the air gaps between the beam delivery system, the patient and the DTC can be assumed to be straight rays, inside the patient they are stochastic and unknown due to electromagnetic (MCS) and nuclear interactions with atomic nuclei. Using the model of MCS by Lynch and Dahl \cite{lynch_approximations_1991} and Gottschalk \cite{gottschalk_multiple_1993}{}, and given the positions and directions $x\hoch{PD}$ from \cref{sec:proton-recon}, the \emph{most likely path} (MLP) can be analytically approximated in a maximum likelihood formalism \cite{schulte_maximum_2008}. The \emph{extended MLP} formalism \cite{krah_comprehensive_2018} also takes uncertainties of the positions and directions into account: this is especially important when the front tracker is omitted, since the beam distribution is modeled using this formalism. Alternatively, a weighted cubic spline is a good approximation of the MLP \cite{Collins-Fekete_theoretical_2017}.  The WEPL of the proton is the integral of the RSP along the estimated path, and has also been reconstructed (as described in \cref{sec:proton-recon}) as $x\hoch{W}$. This leads to a linear system of equations for the list $y$ of all voxels of the RSP image,
\begin{equation}\label{eq:linear-system}
A_{x\hoch{PD}} y = x\hoch{W}.
\end{equation}
The entry $(i,j)$ of the matrix $A=A_{x\hoch{PD}}$ stores how much the RSP at voxel $j$ influences the WEPL of proton $i$; this value is related to the length of intersection of the proton's path and the voxel's volume, and thus depends on $x\hoch{PD}$. As each proton passes through a minor fraction of all voxel volumes, $A$ is typically sparse. Instead of the approximate length of intersection, a constant chordlength, mean chord length or effective mean chord length \cite{penfold_more_2009} can be used to determine the matrix elements. As an alternative, we also study a ``thick paths'' or ``fuzzy voxels'' approach where points on the path are assumed to influence a wider stencil of surrounding voxels, with a weight that decreases with distance.
  
The matrix $A$ is not a square matrix as the number of protons is independent of the dimensions of the RSP image. Therefore ``solving~\eqref{eq:linear-system}'' is either meant in the least-squares sense, or additional objectives like noise reduction are taken into account via regularization or superiorization \cite{penfold_total_2010}.

The two main computational steps of the MBIR subprocedure, generating the matrix $A_{x\hoch{PD}}$ and solving the system~\eqref{eq:linear-system}, are displayed in \cref{fig:recon}. Several implementations of x-ray MBIR algorithms \cite{biguri_tigre_2016,van_aarle_astra_2015} actually do not store the matrix in memory explicitly, because even in a sparse format it would be to large \cite{biguri_tigre_2016}. %
Rather, they implement \emph{matrix-free solvers} like ART, SIRT, SART, DROP (diagonally relaxed orthogonal projections) \cite{penfold_techniques_2015}{} or LSCG (least squares conjugate gradient). These access the matrix only in specific ways, e.\,g.\ via (possibly transposed) matrix-vector products or calculating norms of all rows. %
The matrix elements can thus be regenerated on-the-fly to perform the specific operation demanded by the solver. In pCT, it is best to structure these operations as \emph{policies} to be applied for each row of $A$: Due to the bent paths of protons, it is more efficient to make steps along a path and detect all the voxels it meets, rather than the other way round. 
Parallel architectures like GPUs can provide a significant speedup for operations whose policies can be executed concurrently for many rows.

\subsection{Derivatives of Algorithms}\label{sec:ad}
\subsubsection{Differentiability}
Any computer program that computes a vector $y\in \RR^m$ of \emph{output variables} based on a vector $x\in\RR^n$ of \emph{input variables} defines a function $f: \RR^n \to \RR^m$, that maps $x$ to $y$. 
Typically, $f$ is differentiable for almost all $x$, because we may think of a computer program as a big composition of elementary functions like $+$, $\cdot$, $\sin$, $\sqrt{\phantom{x}}$, $|\cdot|$, and apply the chain rule.
Possible reasons why $f$ could not be differentiable at a particular $x$ include the following: \begin{itemize}
    \item An elementary function is evaluated at an argument where it is not differentiable, like the abs function $|\cdot|$ at $0$.
    \item An elementary function is evaluated at an argument where it is not even continuous, like rounding at $k+\tfrac{1}{2}$ for integers $k$.
    \item An elementary function is evaluated at an argument where it is not even defined, like division by zero.
   \item A control flow primitive such as {\ttfamily if} or {\ttfamily while} makes a comparison like $a=b$, $a>b$, $a<b$, $a \geq b$ or $a \leq b$ between two expressions $a$, $b$ that actually have the same value for the particular input $x$.
\end{itemize}
\subsubsection{Taylor's Theorem} 
Although differentiability and derivatives are local concepts, they can be used for extrapolation by \emph{Taylor's theorem}. A precise statement \cite{anderson_econ_nodate} in first order is that if $f$ is twice continuously differentiable, for each $\xo$ there are constants $C, D \in \RR$ such that the error of the linearization
\begin{equation} \label{eq:taylor}
f(x) \approx f(\xo) + f'(\xo) \cdot (x-\xo)
\end{equation}
is bounded by $C\cdot |x-\xo |^2$ for all $x$ with $|x-\xo |<D$. The constant $C$ depends on how steep $f'$ is. The Taylor expansion \eqref{eq:taylor} is frequently used as a heuristic even if the differentiability requirements are violated. This is only meaningful if the ``amount'' or ``density'' of the non-differentiable points $x$ listed above is low. To assess the range in which the linearization~\eqref{eq:taylor} is valid, one can plot $f(x)-f(\xo)$ against $|x-\xo |$ and compare to the graph of a proportionality relation. We call it the \emph{linearizability range}.

The two applications discussed in the next \cref{sec:purpose-uq,sec:purpose-opt} rely on \eqref{eq:taylor}.

\subsubsection{Quantification of Uncertainty}\label{sec:purpose-uq}
If the value of an input $x$ is approximated by $\xo$ and the resulting uncertainty of $f(x)$ is sought, $f'(\xo)$ contains the relevant information to measure the amplification of errors: according to the Taylor expansion~\eqref{eq:taylor}, the deviation of $x$ from $\xo$ gives rise to an approximate deviation of $f(x)$ from $f(\xo)$ by $f'(\xo) \cdot (x-\xo)$. 

For small local perturbations, a more specific analysis is possible when we model the uncertainty in the input $x$ using a Gaussian distribution with mean $\xo$ and covariance matrix $\Sigma_x$, and replace $f$ with its linear approximation \eqref{eq:taylor}. For a linear function $f$, the output $f(x)$ is a Gaussian distribution with mean $f(\xo)$ and covariance matrix 
\begin{equation}\label{eq:sigmay}
\Sigma_{f(x)} = f'(\xo) \cdot \Sigma_x \cdot f'(\xo)^T. 
\end{equation}

We are interested in uncertainties of the reconstructed RSP image as a result of the MBIR subprocedure, or the uncertainties of results of further image processing like contour lines \cite{aehle_quantification_2021}. Possible input variables of known uncertainty are the detector output or reconstructed proton paths.
All steps in the pipeline from this input to the output must be differentiated in order to apply \eqref{eq:sigmay}. Since the above input variables are defined after the Monte Carlo subprocedure (see \cref{fig:pipeline-coarse}), the differentiation of this step is not required. %

\begin{figure}
    \centering
    \begin{subfigure}{0.45\textwidth}
        \includegraphics[width=\textwidth]{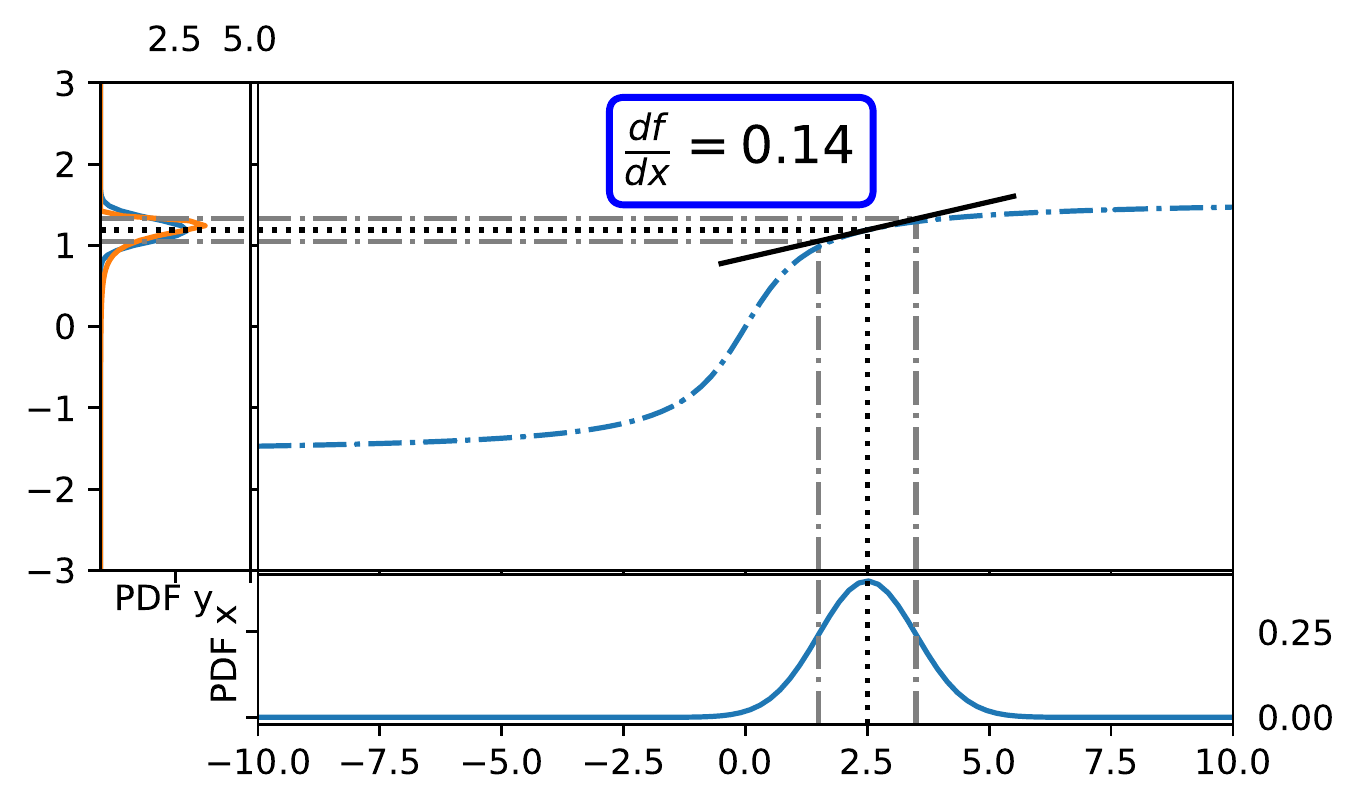}
        \caption{Good approximation, because $f$ is approximatively linear in e.\,g.\ a $\sigma$-envelope around $\xo$.}
        \label{fig:sigmay-good}
    \end{subfigure}\qquad
    \begin{subfigure}{0.45\textwidth}
        \includegraphics[width=\textwidth]{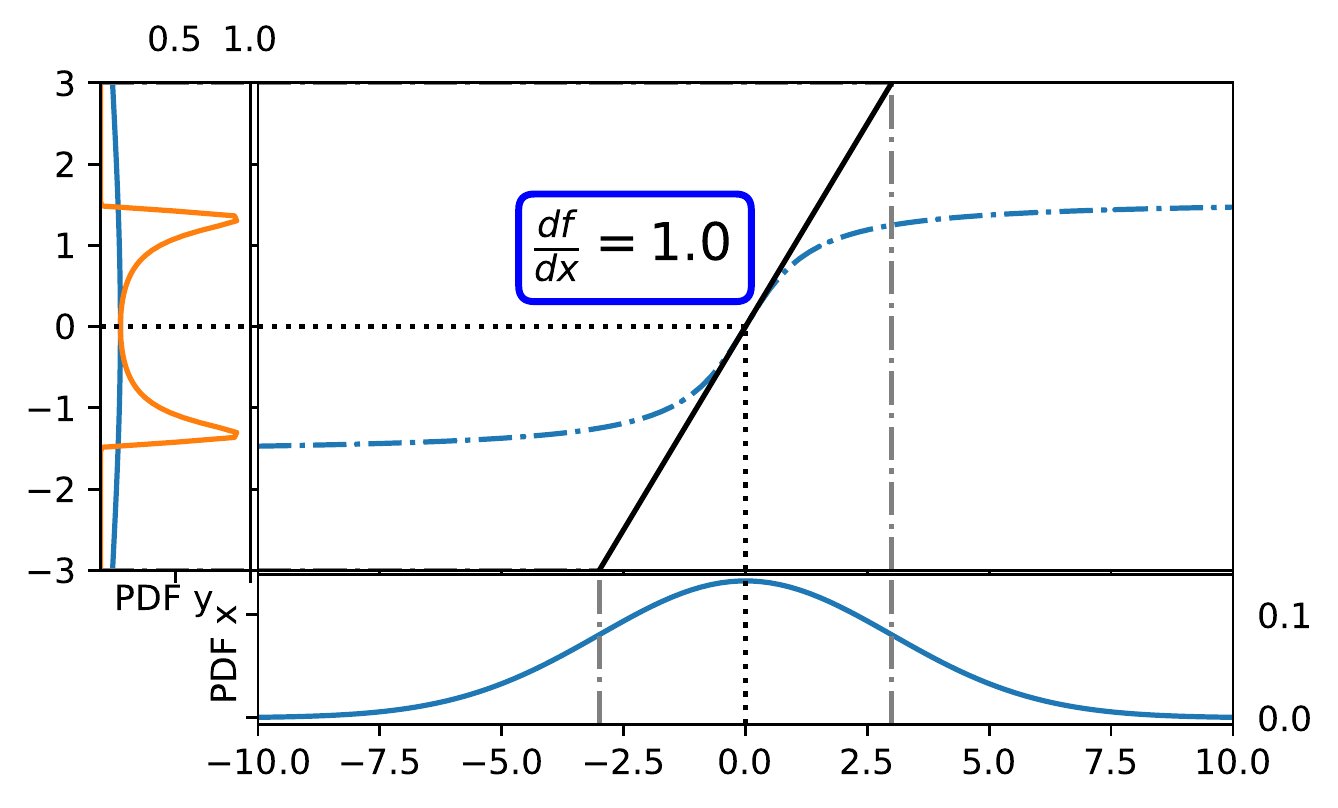}
        \caption{Bad approximation; $f(x)$ is far from being a Gaussian distribution.}
        \label{fig:sigmay-bad}
    \end{subfigure}
    \caption{A function $f$, shown by the dash-dotted plot in the central box, is applied to a Gaussian random variable $x$ whose probability density function (PDF) is shown in the bottom graph. The orange graph in the left box conveys the true distribution of $f(x)$, whereas the normal distribution with variance according to \eqref{eq:sigmay} is shown in blue. Cited from \protect\citet{aehle_quantification_2021}. }
    \label{fig:sigmay}
\end{figure}

Uncertainties in the input should be ``unlikely'' to go beyond the linearizability range. Otherwise, $f$ is not approximated well by its linearization for a significant amount of possible inputs $x$ and \eqref{eq:sigmay} cannot be used, as illustrated in \cref{fig:sigmay}. 

No matter how non-smooth $f$ is, 
Monte Carlo simulation can be used alternatively to propagate a random variable $x$ through a computer program $f$ and estimate the statistical properties of the random variable $f(x)$. The more samples are used, the better the Monte Carlo estimate becomes, but the more run-time must be spent. 

\subsubsection{Gradient-Based Optimization}\label{sec:purpose-opt}
In the case that $f$ has a single output variable ($m=1$), optimization seeks to find a value for $x$ that minimizes (or maximizes) this \emph{objective function}. The idea behind gradient-based optimization is that as the gradient points into the \emph{direction of steepest ascent} according to \eqref{eq:taylor}, shifting $x$ in the opposite direction should make $f(x)$ smaller. Neural network training algorithms like \emph{ADAM} also belong to this category, with the training error of the neural network as the objective function.

In order to perform an end-to-end  optimization of the detector parameters to minimize an objective function based on the reconstructed RSP, all parts of the pipeline in \cref{fig:pipeline-coarse} would have to be differentiated, including the Monte Carlo simulation.

Optimization algorithms can only find \emph{local minima}, which are optimal among ``similar'' designs but possibly worse than the global minimum. Lack of differentiability and consequently, lack of explicit gradients preclude the guarantee that a gradient-based optimization algorithm can converge to a local minimum. 
Instead of using the gradient to find a descent direction, one could alternatively iterate through all coordinate directions, use random choices or a surrogate model. Many \emph{derivative-free optimization} algorithms have been proposed in the literature \cite{larson_derivative-free_2019}. However, few of them are able to deal with 300 unknowns or more \cite{rios_derivative-free_2013}.

\subsubsection{Algorithmic Differentiation}\label{sec:diff-algorithmic}

The classical ways to obtain derivatives of a computer-implemented function $f$, required for the applications described in \cref{sec:purpose-uq,sec:purpose-opt}, are \emph{analytical} (using differentiation rules, possible only for simple programs) or \emph{numerical} (using difference quotients, inexact). Ideally, \emph{algorithmic} differentiation combines their respective advantages being exact and easy applicable. AD tools facilitate the application of AD to an existing codebase; specifically, \emph{operator overloading} type AD tools intercept floating-point arithmetic operators and math functions, and insert AD logic that keeps track of derivatives with respect to input variables (\emph{forward mode}), or records an arithmetic evaluation tree (\emph{reverse mode}). As examples for such tools, we may cite  ADOL-C \cite{Walther2012Gsw}{}, CoDiPack \cite{SaAlGauTOMS2019}, the autograd tool \cite{maclaurin2015autograd} used by PyTorch \cite{NEURIPS2019_9015}, and the internal AD tool of TensorFlow \cite{tensorflow2015-whitepaper}. The machine-code based tool Derivgrind \cite{aehle_forward-mode_2022,aehle_reverse-mode_2022} may offer a chance to integrate AD into cross-language and partially closed-source software projects.
For an overview of tools and applications, visit \url{https://www.autodiff.org}.

While these tools can often be applied ``blindly'' to any computer program to obtain algorithmic derivatives in an ``automatic'' fashion, further program-specific adaptations might be necessary. For example, the new datatype of an operator overloading tool might break assumptions on the size or format of the floating-point type that were hard-coded in the original program. Concerning complex simulations, techniques like \emph{checkpointing} \cite{hutchison_data-flow_2006,naumann_adjoint_2018} or \emph{reverse accumulation} can reduce the memory consumption of the tape in reverse mode but require manual modifications of the primal program. 

\begin{figure}
\centering
\begin{subfigure}{0.3\textwidth}
\begin{tikzpicture}
\begin{axis}[xlabel={$x$},ylabel={$y$},ticks=none,height=4.5cm,width=4.5cm,xmin=-1.2,xmax=1.2,ymin=-0.2,ymax=1.2]
\addplot[domain=-1:1,samples=200,black,very thick]  {x*x};
\addplot[blue,domain=-1:1,samples=200]  {x*x - 0.1*x*x*x + 0.1*sin(200*x) - 0.1}; 
\end{axis}
\end{tikzpicture}
\caption{~~\,}
\label{fig:badapproxderiv-a}
\end{subfigure}
\begin{subfigure}{0.3\textwidth}
\begin{tikzpicture}
\begin{axis}[xlabel={$x$},ylabel={$y$},ticks=none,height=4.5cm,width=4.5cm,xmin=-1.2,xmax=1.2,ymin=-0.2,ymax=1.2]
\addplot[domain=-1:1,samples=200,black,very thick]  {x*x};
\addplot[blue,domain=-1:1,samples=200]  {0.05+floor(x*x*10)/10 }; 
\end{axis}
\end{tikzpicture}
\caption{~~\,}
\label{fig:badapproxderiv-b}
\end{subfigure}
\begin{subfigure}{0.3\textwidth}
\begin{tikzpicture}
\begin{axis}[xlabel={$x$},ylabel={$y$},ticks=none,height=4.5cm,width=4.5cm,xmin=-1.2,xmax=1.2,ymin=-0.2,ymax=1.2]
\addplot[domain=-1:1,samples=200,black,very thick]  {x*x};
\addplot[blue,domain=-1:1,samples=600]  {x*x  + 0.05*sin(20000*x) + 0.05*cos(31400*x*x)}; 
\end{axis}
\end{tikzpicture}
\caption{~~\,}
\label{fig:badapproxderiv-c}
\end{subfigure}
\caption{Three value-wisely good approximations (blue thin line) to a smooth function (black thick line). In (a), the derivative of the approximation is also a good approximation for the derivative of the smooth function. Such a statement can however be wrong, e.\,g.\ if there are jumps as in (b) or noise as in (c).}
\label{fig:badapproxderiv}
\end{figure}
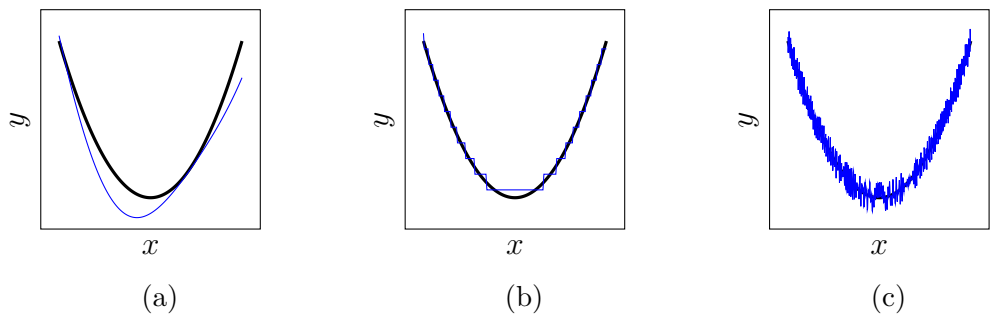

Another major reason for revisiting the primal code is given by the fact that it is usually only an approximation of the real-world process. Good function approximation does not guarantee that the corresponding derivatives are also well-approximated \cite{sirkes_finite_1997}. To illustrate this, \cref{fig:badapproxderiv} shows three value-wisely good approximations to a smooth function. In \cref{fig:badapproxderiv-b}, the derivative of the approximation is zero everywhere except where the approximation jumps. This kind of behaviour could be the consequence of intermediate rounding steps. In \cref{fig:badapproxderiv-c}, a low-magnitude but high-frequency error adds high-magnitude noise to the derivative. %
In both cases, the exact AD derivative of the approximation is entirely unrelated to the derivative of the real-world function, and therefore cannot be of any use to the propagation of uncertainties through it, or its optimization. Adaptations of the computer program might be necessary to ensure that it also a good approximation derivative-wise, as in \cref{fig:badapproxderiv-a}.

\subsection{Numerical Checks of Algorithmic Differentiability}\label{sec:methodology}
In the work presented in this article, we determined the potential of employing AD for gradient-based optimization or UQ of the pCT reconstruction pipeline. 

Our main methodology was based on plots that show the dependency of a single output variable $f(x)$ with respect to a single input variable $x$ for three representative substeps of the pipeline, keeping all the other inputs fixed. More details on the three setups can be found in \crefrange{sec:methodology-gate}{sec:methodology-mbir}.

From a plot of $f(x)$ with a linearly scaled abscissa, we identified whether $f$ had isolated discontinuities (``jumps'') or noise. Differentiability (from one side) at a particular $\xo$ could be verified with a log-log plot of $|f(x)-f(\xo)|$ with respect to $x-\xo$. Inside the linearizability range, it should look like the log-log plot of a proportionality relation, i.\,e.\ a straight line with slope~1. Deviations for very small $|x-\xo|$ can often be attributed to floating-point imprecisions and have no implications on the differentiability. 

If we choose several points $\xo$ without special considerations in mind, and $f$ turns out to be differentiable at all of them, we take this as an indicator that the function is differentiable ``almost everywhere''. This means that the function may have e.\,g.\ jumps or kinks, but those are unlikely to be encountered with generic input.

\subsubsection{Setup for GATE}\label{sec:methodology-gate}
We used GATE~v9.1 to simulate a single proton of initial energy $x$ around $\xo = \SI{230}{\mega\eV}$ (as well as other arbitrary values \SI{220.123}{\mega\eV}, \SI{229}{\mega\eV}, \SI{231}{\mega\eV}, \SI{240}{\mega\eV}, \SI{256.987}{\mega\eV}) passing through a head phantom \cite{GIACOMETTI2017182} and several layers of the DTC. Four output variables were extracted from the ROOT file produced by GATE: the energy depositions $f_{\text{E},1}(x)$, $f_{\text{E},2}(x)$, and a position coordinate $f_{\text{pos},1}(x)$, $f_{\text{pos},2}(x)$, in the first and second tracking layer of the DTC, respectively. The seed of the RNG has been kept constant.

\subsubsection{Setup for Tracking}\label{sec:methodology-tracking}

We analyzed the percentage of correctly reconstructed tracks of a track-following scheme implemented by \citet{pettersen_design_2019,pettersen_proton_2020}. During the reconstruction, a threshold determined the maximal accumulated angular deflections allowed for continued reconstruction. This threshold was then identified as the input variable to test for differentiability. A batch of \num{10000} tracks was used for this purpose, where correctly reconstructed tracks were identified on the basis of their Bragg peak position relative to the MC truth (\SI{\pm 2}{\cm}) \cite{pettersen_investigating_2021}.

\subsubsection{Setup for MBIR}\label{sec:methodology-mbir}
\begin{figure}
\centering
\begin{subfigure}[t]{0.44\textwidth}
\centering
\includegraphics[width=0.9\textwidth]{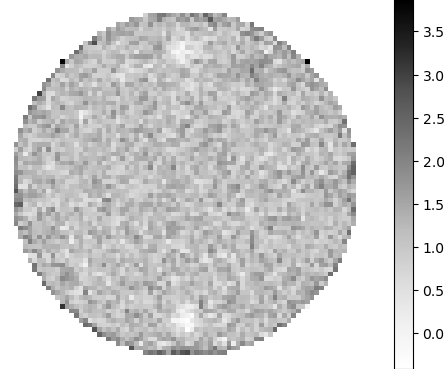}
\caption{DROP, 0.25 million protons.}
\label{fig:mbir-solutions-drop}
\end{subfigure}
\begin{subfigure}[t]{0.44\textwidth}
\centering
\includegraphics[width=0.9\textwidth]{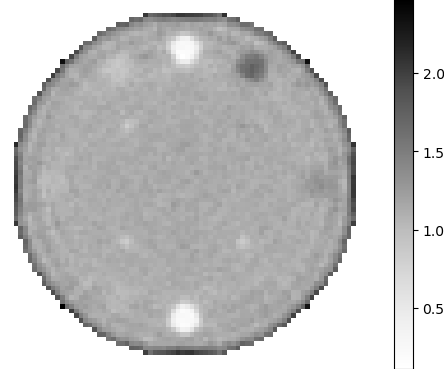}
\caption{DROP, 25 million protons.}
\label{fig:mbir-solutions-drop-fine}
\end{subfigure}
\\
\begin{subfigure}[t]{0.44\textwidth}
\centering
\includegraphics[width=0.9\textwidth]{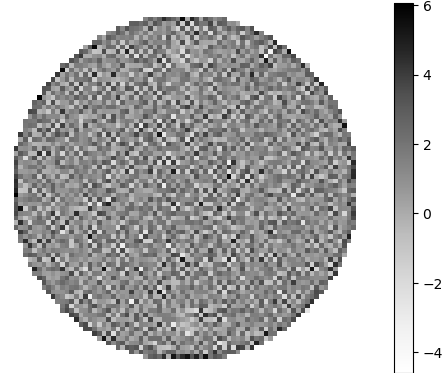}
\caption{LSCG, 0.25 million protons.}
\label{fig:mbir-solutions-lscg}
\end{subfigure}
\begin{subfigure}[t]{0.44\textwidth}
\centering
\includegraphics[width=0.9\textwidth]{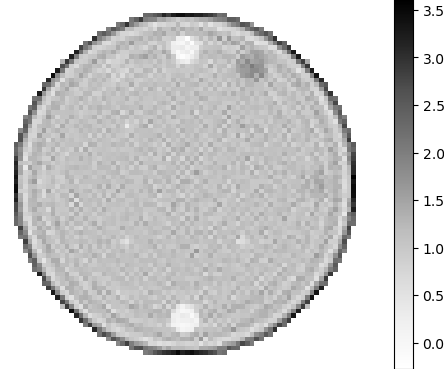}
\caption{LSCG, 25 million protons.}
\label{fig:mbir-solutions-lscg-fine}
\end{subfigure}
\caption{Central slices of reconstructed RSP images.}
\label{fig:mbir-solutions}
\end{figure}

We considered a setup with about 250\,000 proton histories from a GATE simulation of the CTP404 phantom \cite{catphan-manual} (an epoxy cylinder of radius \SI{75}{\milli\meter} and height \SI{25}{\milli\meter}, containing cylindric inserts of various other materials), and reconstructed an RSP image with 5 slices (\SI{8}{\milli\meter} thick) of $80 \times 80$ voxels ($\SI{2}{\milli\meter} \times \SI{2}{\milli\meter}$) using our own prototypical implementations of DROP and LSCG using C{\ttfamily++}, CUDA and Python. 
The setup was small enough to allow for many repetitive evaluations with modifications in a single input variable within a reasonable computing time. Both DROP (with a relaxation factor of $0.1$) and LSCG seem to converge for the over-determined linear system \eqref{eq:linear-system}.   \Cref{fig:mbir-solutions-drop,fig:mbir-solutions-lscg} show the central slice of the reconstructed RSP image after 400~iterations of either solver. 

In terms of the root mean square error of the linear system \eqref{eq:linear-system}, the more noisy LSCG solution is better than the DROP solution by about \SI{7}{\percent}. In fact, LSCG seems not to be widespread in CT image reconstruction, but we included it in our study for the purpose of comparison.

The solutions in \cref{fig:mbir-solutions-drop-fine,fig:mbir-solutions-lscg-fine} were reconstructed from about 25~million proton histories; they indicate that the overall bad image quality of \cref{fig:mbir-solutions-drop,fig:mbir-solutions-lscg} is an artifact of the low number of proton histories. It should be irrelevant for the qualitative statement on differentiability.

We separately considered two input variables: the WEPL (i.\,e.\ a component of the right-hand side $x\hoch{W}$ of \eqref{eq:linear-system}) and a coordinate of the beam spot position of one particular proton history (i.\,e.\ a component of $x\hoch{PD}$, influencing the matrix in \eqref{eq:linear-system}).

To study the dependency on the WEPL, we applied the DROP algorithm with a relaxation factor of 0.1 as well as the LSCG algorithm, using a mean chord length approach to compute the matrix elements. We observed the reconstructed RSP of a voxel that the input proton history passed through. For DROP, geometric information was used by zeroing voxels outside a cylindrical hull of radius \SI{75}{\milli\meter} after each iteration.

To study the dependency on the position coordinate, we applied the DROP algorithm with a relaxation factor of 0.1, using a mean chord length approach as well as a fuzzy voxels approach to compute the matrix elements. In the fuzzy voxels approach, voxels in a $3\times 3$ neighbourhood around points on the path received a weight that decreased exponentially with the distance of the center of the voxel, across all slices. We observed the reconstructed RSP of a voxel that the unmodified input proton history passed through.

\subsubsection{Recording of RNG Calls}\label{sec:methodology-rngtraces}
To understand the cause of jumps observed in the setup of \cref{sec:methodology-gate}, we additionally performed the following analysis. After identifying the precise location of the jump via bisection, we used a debugger to output the backtrace of every call to the RNG. After masking floating-point numbers and pointers, we obtained a medium-granular record of the control flow in the program for the particular input used to run it. We produced four of these records in close proximity to one particular jump, two on each side.

\section{Results}\label{sec:results}

\subsection{Monte Carlo Subprocedure}\label{sec:results-mc}

\begin{figure}
\centering
\begin{tikzpicture}
\begin{axis}[xtick={229.6,230,230.4}, xticklabels={229.6,230,230.4}, extra x tick style={grid=major}, xmajorgrids, xlabel={$x$ in \SI{1}{\mega\eV}},ylabel style={align=center}, ylabel={energy deposition \\ in \SI{1}{\mega\eV}},height=5cm]
\addplot[gray,mark=*,mark size=0.8,only marks] table[x index=0, y index=3] {images/around_230_head/yy};
\addplot[black,mark=*,mark size=0.4,only marks] table[x index=0, y index=1] {images/around_230_head/yy};
\end{axis}
\end{tikzpicture}
\begin{tikzpicture}
\pgfplotsset{set layers}
\begin{axis}[xtick={229.6,230,230.4}, xticklabels={229.6,230,230.4}, extra x tick style={grid=major}, xmajorgrids, xlabel={$x$ in \SI{1}{\mega\eV}},ylabel style={align=center},ylabel={position coord.\\ in \SI{1}{\milli\meter}},height=5cm]
\addplot[gray,mark=*,mark size=0.8,only marks] table[x index=0, y index=4] {images/around_230_head/yy};
\addplot[black,mark=*,mark size=0.4,only marks] table[x index=0, y index=2] {images/around_230_head/yy};
\end{axis}
\end{tikzpicture}
\\ 
\centering
\begin{tikzpicture}
\begin{axis}[xmin=230.1055,xmax=230.1075,xtick={230.106,230.107}, xticklabels={230.106,230.107}, extra x tick style={grid=major}, xmajorgrids, xlabel={$x$ in \SI{1}{\mega\eV}},ylabel style={align=center}, ylabel={energy deposition \\ in \SI{1}{\mega\eV}},height=5cm]
\addplot[gray,mark=*,mark size=0.8,only marks] table[x index=0, y index=3] {images/around_230_head/yy-zoom};
\addplot[black,mark=*,mark size=0.4,only marks] table[x index=0, y index=1] {images/around_230_head/yy-zoom};
\end{axis}
\end{tikzpicture}
\begin{tikzpicture}
\pgfplotsset{set layers}
\begin{axis}[xmin=230.1055,xmax=230.1075,xtick={230.106,230.107}, xticklabels={230.106,230.107}, extra x tick style={grid=major}, xmajorgrids, xlabel={$x$ in \SI{1}{\mega\eV}},ylabel style={align=center},ylabel={position coord.\\ in \SI{1}{\milli\meter}},height=5cm]
\addplot[gray,mark=*,mark size=0.8,only marks] table[x index=0, y index=4] {images/around_230_head/yy-zoom};
\addplot[black,mark=*,mark size=0.4,only marks] table[x index=0, y index=2] {images/around_230_head/yy-zoom};
\end{axis}
\end{tikzpicture}
\caption{Numerical check whether GATE is well-linearizable: The energy deposition $f_{\mathrm E, j}(x)$ (left) and a position coordinate $f_{\text{pos},j}(x)$ (right) in the first ($j=1$, black) and second ($j=2$, gray) tracking layer are plotted against the beam energy $x$ in a wide (top) and narrow (bottom) interval. }
\label{fig:gate-diff-study-around_230}
\end{figure}
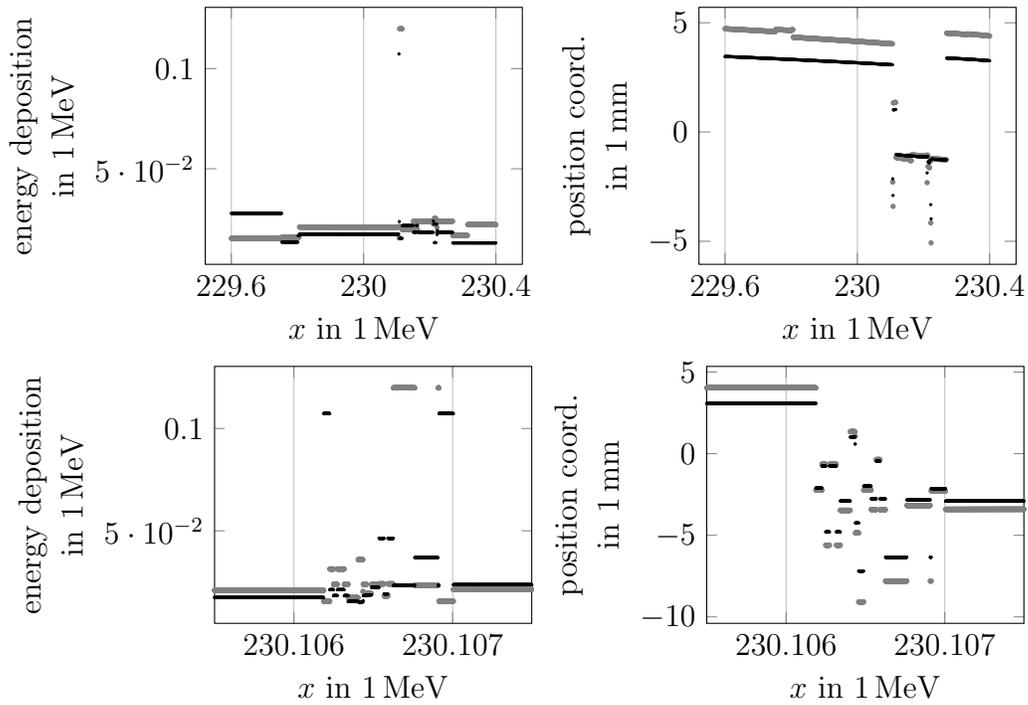

In the top row of \cref{fig:gate-diff-study-around_230} for the GATE setup of \cref{sec:methodology-gate}, $f_{\text{E},1}$, $f_{\text{E},2}$, $f_{\text{pos},1}$, $f_{\text{pos},2}$ appear as piecewise differentiable functions with around one jump per \SI{0.1}{\mega\eV}. %

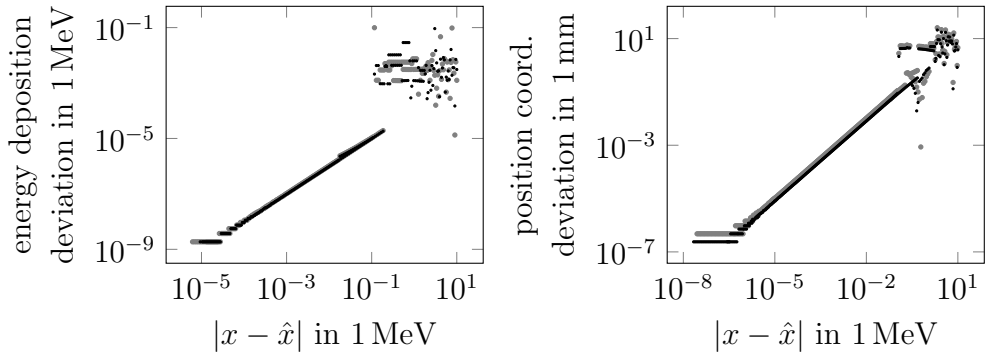
\begin{figure}
\centering
\begin{tikzpicture}
\begin{axis}[ymode=log,xmode=log,xlabel={$|x-\xo|$ in \SI{1}{\mega\eV}},ylabel style={align=center}, ylabel={energy deposition \\ deviation in \SI{1}{\mega\eV}},height=5cm]
\addplot[gray,mark=*,mark size=0.8,only marks] table[x index=0, y index=3] 
{images/around_230_head/absdiff-230};
\addplot[black,mark=*,mark size=0.4,only marks] table[x index=0, y index=1] 
{images/around_230_head/absdiff-230};
\end{axis}
\end{tikzpicture}
\begin{tikzpicture}
\begin{axis}[ymode=log,xmode=log,xlabel={$|x-\xo|$ in \SI{1}{\mega\eV}},ylabel style={align=center}, ylabel={position coord.\\ deviation in \SI{1}{\milli\meter}},height=5cm]
\addplot[gray,mark=*,mark size=0.8,only marks] table[x index=0, y index=4] 
{images/around_230_head/absdiff-230};
\addplot[black,mark=*,mark size=0.4,only marks] table[x index=0, y index=2] 
{images/around_230_head/absdiff-230};
\end{axis}
\end{tikzpicture}
\caption{These are the same plots as in \cref{fig:gate-diff-study-around_230}, with a different axis range and scale. As $|f_{\cdot}(x)-f_{\cdot}(\xo)|$ is about proportional to $x-\xo$ if $x$ is close enough to $\xo=\SI{230}{\mega\eV}$, the four functions are differentiable at this point. }

\label{fig:gate-diff-study-log_230_240}
\end{figure}

For low perturbations $x-\xo$ of the beam energy around $\xo = \SI{230}{\mega\eV}$, the log-log plot in \cref{fig:gate-diff-study-log_230_240} shows a straight line with slope $1$, indicating that the four functions were differentiable at $\xo$.  After some threshold perturbation given by the distance to the next discontinuity, the approximation error of the energy depositions $f_{\text{E},1}$, $f_{\text{E},2}$ rose suddenly. Analogous observations were made for the other test values of $\xo$ listed in \cref{sec:methodology-gate}.

Zooming in around the jump at \SI{230.106}{\mega\eV} (bottom row of \cref{fig:gate-diff-study-around_230}), we observe that it is actually a cluster of many discontinuities. We further investigated two discontinuities using masked records of backtraces of RNG calls (\cref{sec:methodology-rngtraces}). Choosing four input values close to \SI{230.106187089}{\mega\eV} by appending a digit 4, 5, 6 or 7 to this decimal representation, we obtained two inputs on each side of a discontinuity. The records for inputs on the same side agreed. When inputs from both sides were used, the control flows differed at some point because the physical interaction length (PIL) compared differently to the current physical step size, leading to different processes being selected as PIL ``winners''. Similarly, the records on both sides of the discontinuity at \SI{230.106225175}{\mega\eV} started to deviate at a comparison between step lengths for ``soft'' and ``hard'' scattering in the Wentzel VI model \cite{FERNANDEZVAREA1993447,urban,Ivanchenko_2010}. The comparison was sensitive to perturbations of the input because the step lengths happened to take values very close to each other. The result of the comparison controlled an early exit from a loop that calls the RNG in each iteration. 

\subsection{Track Reconstruction}\label{sec:results-tracking}

\begin{figure}
\centering
\hfill
\begin{subfigure}[t]{0.30\textwidth}
\centering
\begin{tikzpicture}
\begin{axis}[xlabel={$x$},ylabel style={align=center},ylabel={accuracy in \%},width=0.9\textwidth]
\pgfplotstableset{
    create on use/yy/.style={create col/copy column from table={images/tracking/tracking-yy}{0}}
}
\addplot[black,mark=*,mark size=0.5pt,only marks] table[y=yy] {images/tracking/tracking-xx};
\end{axis}
\end{tikzpicture}
\caption{Full range, showing potential for optimization.}
\label{fig:tracking-diff-study-a}
\end{subfigure}
\hfill
\begin{subfigure}[t]{0.30\textwidth}
\centering
\begin{tikzpicture}
\begin{axis}[xlabel={$x$},ylabel style={align=center}, ylabel={accuracy in \%},width=0.9\textwidth]
\pgfplotstableset{
    create on use/yy/.style={create col/copy column from table={images/tracking/tracking-yy-zoom}{0}}
}
\addplot[black,mark=*,mark size=0.5pt,only marks] table[y=yy] {images/tracking/tracking-xx-zoom};
\end{axis}
\end{tikzpicture}
\caption{Zooming in, we can see noise.}
\label{fig:tracking-diff-study-b}
\end{subfigure}
\hfill
\begin{subfigure}[t]{0.30\textwidth}
\centering
\begin{tikzpicture}
\begin{axis}[xlabel={$x$},ylabel style={align=center}, ylabel={accuracy in \%},width=0.9\textwidth]
\pgfplotstableset{
    create on use/yy/.style={create col/copy column from table={images/tracking/tracking-yy-more-zoom}{0}}
}
\addplot[black,mark=*,mark size=0.5pt,only marks] table[y=yy] {images/tracking/tracking-xx-more-zoom};
\end{axis}
\end{tikzpicture}
\caption{Zooming in even further, we can see steps.}
\label{fig:tracking-diff-study-c}
\end{subfigure}
\hfill
\caption{Dependency of the tracking accuracy in \% on the vertical axis w.\,r.\,t.\ a threshold $x$ used by the track-following scheme. \Crefrange{fig:tracking-diff-study-a}{fig:tracking-diff-study-c} only differ in the range of the horizontal axis. }
\label{fig:tracking-diff-study}
\end{figure}
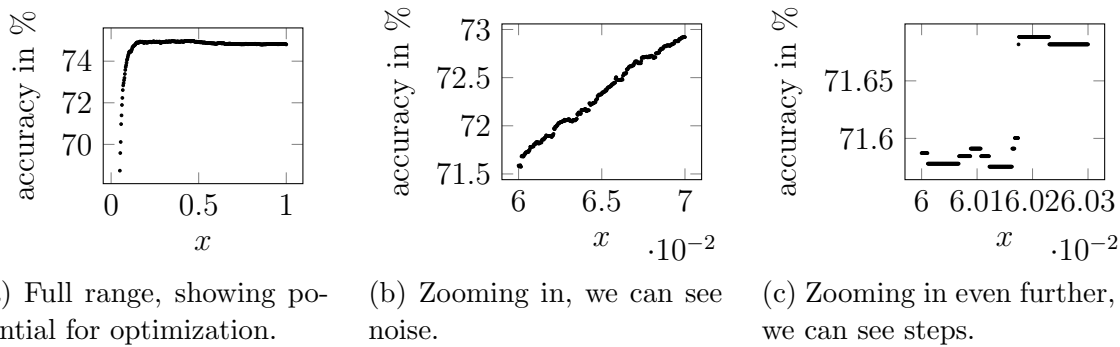

The global behaviour in \cref{fig:tracking-diff-study-a} shows that the investigated input parameter offers potential for optimization of the tracking accuracy (see \cref{sec:methodology-tracking}) by choosing it sufficiently large.

\Cref{fig:tracking-diff-study-b} deals with medium-sized modifications. The high number of steps and the noisy behaviour of the plot in \cref{fig:tracking-diff-study-b} indicate that the code uses non-differentiable operations very frequently, so probably linearizability ranges of other output variables are very small as well.

\Cref{fig:tracking-diff-study-c} displays the effect of very small modifications of the input parameter. As the percentage of correctly reconstructed tracks is an inherently discrete quantity, we expect to see steps here, instead of a gradual transition.

\subsection{RSP Reconstruction}\label{sec:results-mbir}

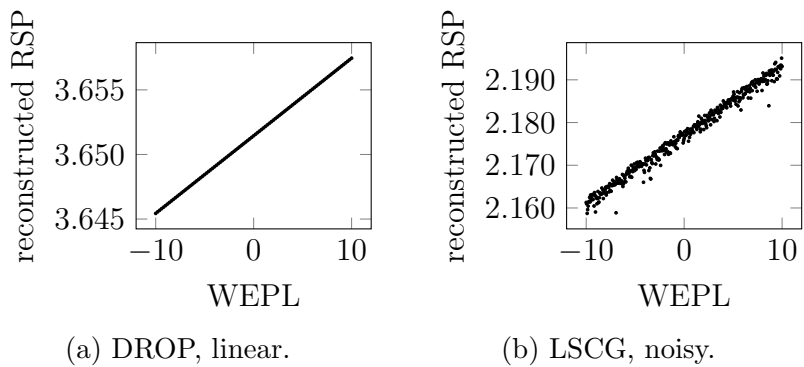
\begin{figure}
\centering
\begin{subfigure}[t]{0.3\textwidth}
\begin{tikzpicture}
\begin{axis}[xlabel={WEPL},ylabel style={align=center}, ylabel={reconstructed RSP},width=\textwidth,
y tick label style={/pgf/number format/.cd,fixed,fixed zerofill,precision=3,/tikz/.cd},]
\addplot[black,mark=*,mark options={mark size=0.5pt,solid},only marks] table {images/modify-wepl-drop/xy400};
\end{axis}
\end{tikzpicture}
\caption{DROP, linear.}
\label{fig:ad-modify-wepl-drop}
\end{subfigure}
\qquad 
\begin{subfigure}[t]{0.3\textwidth}
\begin{tikzpicture}
\begin{axis}[xlabel={WEPL},ylabel style={align=center}, ylabel={reconstructed RSP},width=\textwidth,
y tick label style={/pgf/number format/.cd,fixed,fixed zerofill,precision=3,/tikz/.cd},]
\addplot[black,mark=*,mark options={mark size=0.5pt,solid},only marks] table {images/modify-wepl-lscg/xy400};
\end{axis}
\end{tikzpicture}
\caption{LSCG, noisy.}
\label{fig:ad-modify-wepl-lscg}
\end{subfigure}
    \caption{Dependency of the reconstructed RSP at a particular voxel on the WEPL of a particular track. While a negative WEPL would never be measured, it can serve as an input to both MBIR algorithms without problems.}
    \label{fig:ad-modify-wepl}
\end{figure}

As shown in \cref{fig:ad-modify-wepl-drop}, the RSP computed by the DROP algorithm depended linearly on the WEPL in the setup of \cref{sec:methodology-mbir}. This statement is accurate up to floating-point accuracy. The non-linear LSCG algorithm introduced noise, as reported in  \cref{fig:ad-modify-wepl-lscg}. 

When a beam spot coordinate of the first track was modified, changes in the set of voxels traversed by the MLP lead to jumps in the DROP-reconstructed RSP, as shown in \cref{fig:ad-drop-track-coordinate-global}. Between these discontinuities, the graph is almost linear (\cref{fig:ad-drop-track-coordinate-step}), and the log-log plot in \cref{fig:ad-drop-track-coordinate-step-log} numerically verifies that it was differentiable at $\hat x = -63.55$. Tangents at this point were almost horizontal in \cref{fig:ad-drop-track-coordinate-global}, so the reconstructed RSP changed much more via jumps than it did in a differentiable manner. 

\begin{figure}
\centering
\begin{subfigure}[t]{\textwidth}
\begin{tikzpicture}
\begin{axis}[xlabel={beam spot coordinate $x$ of first track},ylabel style={align=center},ylabel={reconstructed RSP $f(x)$},height=5cm,width=0.9\textwidth,legend pos=north west, xmin=-69, xmax=-49,%
extra x ticks={-67.475, -65.525, -63.575, -61.575, -59.625, -59.475, -58.875, -58.525, -57.875, -57.625, -57.125, -56.775, -55.675, -55.625, -54.825, -54.425, -53.675, -53.575, -53.125, -52.175, -51.725, -51.675, -50.675, -50.125, -49.725, -49.525}, extra x tick labels={},extra tick style={grid=major}, %
]
\draw[] (axis cs:-63.575,-1e-4,0)--(axis cs:-63.575,1e-4) (axis cs:-57.125,1e-4)--(axis cs:-57.125,-1e-4) ;
\addplot[black!40,mark=*,mark options={mark size=0.5pt,solid},only marks] table {images/modify-position-pathprojectorfuzzy/xy100};
\addplot[black!60,mark=*,mark options={mark size=0.5pt,solid},only marks] table {images/modify-position-pathprojectorfuzzy/xy200};
\addplot[black!80,mark=*,mark options={mark size=0.5pt,solid},only marks] table {images/modify-position-pathprojectorfuzzy/xy300};
\addplot[black,mark=*,mark options={mark size=0.5pt,solid},only marks] table {images/modify-position-pathprojectorfuzzy/xy400};
\addplot [black!40, dashed, domain=-65:-61.1, smooth] {2.129470626483608e-06-4.404664753334815e-08*(x+63.55)};
\addplot [black!60, dashed, domain=-65:-61.1, smooth] {4.242847305605582e-06-8.804816341338604e-08*(x+63.55)};
\addplot [black!80, dashed, domain=-65:-61.1, smooth] {6.340537572698287e-06-1.3200274498023652e-07*(x+63.55)};
\addplot [black, dashed, domain=-65:-61.1, smooth] {8.422929910547277e-06-1.7590858957402238e-07*(x+63.55)};
\addlegendentry{100 iter.}
\addlegendentry{200 iter.}
\addlegendentry{300 iter.}
\addlegendentry{400 iter.}
\end{axis}
\end{tikzpicture}
\caption{Global perspective for various numbers of DROP steps. At $\hat x=-63.55$, $f(x)$ is differentiable as verified in \cref{fig:fuzzy-ad-drop-track-coordinate-step-log} (in the case of 400~iterations), and the dashed tangent lines at $\hat x$ reflect the global behaviour very well. }
\label{fig:fuzzy-ad-drop-track-coordinate-global}
\end{subfigure} \\[0.3cm]
\begin{subfigure}[t]{0.4\textwidth}
\begin{tikzpicture}
\begin{axis}[xlabel={$x$},ylabel style={align=center},ylabel={$f(x)$},height=5cm,width=0.9\textwidth,%
xmin=-63.575,xmax=-61.575,
y tick label style={/pgf/number format/.cd,fixed,fixed zerofill,precision=3,/tikz/.cd},]
\addplot[black,mark=*,mark options={mark size=0.5pt,solid},only marks] table {images/modify-position-pathprojectorfuzzy/xy400-single-step};
\end{axis}
\end{tikzpicture}
\caption{Zoom into on of the steps of \cref{fig:fuzzy-ad-drop-track-coordinate-global}, with 400 iterations.}
\label{fig:fuzzy-ad-drop-track-coordinate-step}
\end{subfigure}
\qquad
\begin{subfigure}[t]{0.4\textwidth}
\begin{tikzpicture}
\begin{axis}[xlabel={$x-\hat x$},ylabel style={align=center},ylabel={$|f(x)-f(\hat x)|$},height=5cm,width=0.9\textwidth,%
ymode=log,xmode=log,]
\addplot[black,mark=*,mark options={mark size=0.5pt,solid},only marks] table  {images/modify-position-pathprojectorfuzzy-log/xy400-diff};
\end{axis}
\end{tikzpicture}
\caption{Logarithmic plot of the numerator and denominator of the difference quotient in the same range as \cref{fig:fuzzy-ad-drop-track-coordinate-step}.}
\label{fig:fuzzy-ad-drop-track-coordinate-step-log}
\end{subfigure}

\caption{Dependency of a particular voxel's RSP on a particular track's beam spot coordinate. Elements of the system matrix $A$ were calculated using the a ``fuzzy voxels'' approach.}
\label{fig:fuzzy-ad-drop-track-coordinate}
\end{figure}
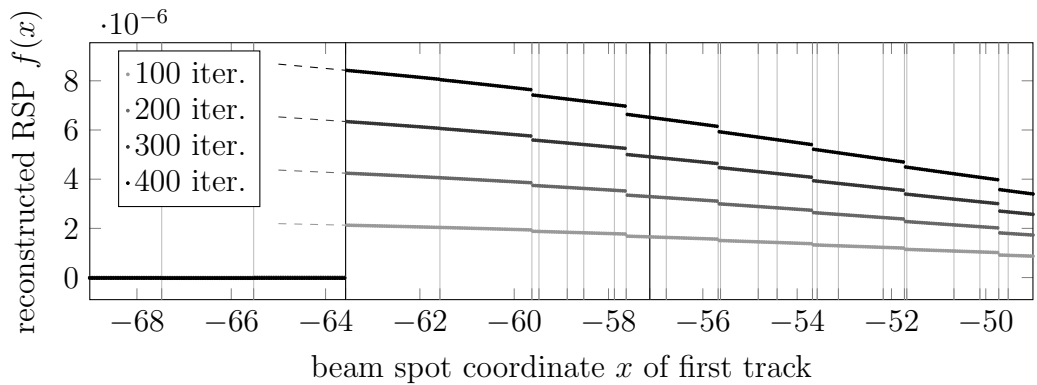
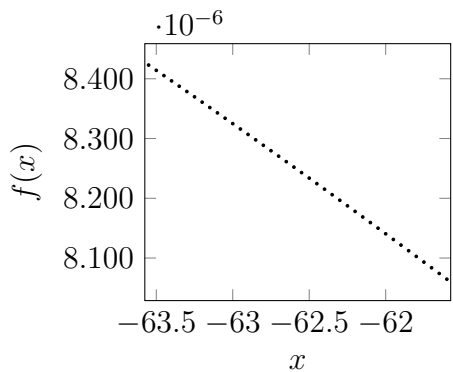
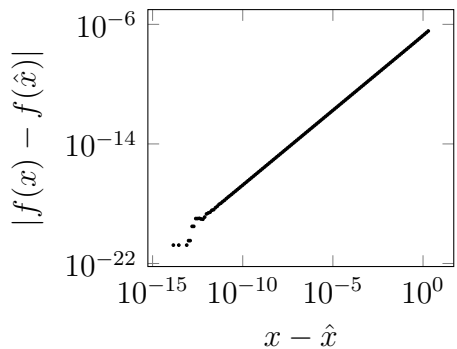

\Cref{fig:fuzzy-ad-drop-track-coordinate} corresponds to \cref{fig:ad-drop-track-coordinate}, but used a fuzzy voxels approach to compute the matrix elements. The graph is still discontinuous wherever the set of voxels traversed by the MLP changes. However, the jumps were much smaller and in between, the functions changed significantly in an almost linear (\cref{fig:fuzzy-ad-drop-track-coordinate-step}) and differentiable (\cref{fig:fuzzy-ad-drop-track-coordinate-step-log}) manner.

\begin{figure}
\centering
\begin{subfigure}[t]{\textwidth}
\begin{tikzpicture}
\begin{axis}[xlabel={beam spot coordinate $x$ of first track},ylabel style={align=center},ylabel={reconstructed RSP $f(x)$},height=5cm,width=0.9\textwidth,legend pos=north west, xmin=-69, xmax=-49,%
extra x ticks={-67.475, -65.525, -63.575, -61.575, -59.625, -59.475, -58.875, -58.525, -57.875, -57.625, -57.125, -56.775, -55.675, -55.625, -54.825, -54.425, -53.675, -53.575, -53.125, -52.175, -51.725, -51.675, -50.675, -50.125, -49.725, -49.525}, extra x tick labels={},extra tick style={grid=major}, %
]
\draw[] (axis cs:-63.575,-1e-3,0)--(axis cs:-63.575,1e-3) (axis cs:-57.125,1e-3)--(axis cs:-57.125,-1e-3) ;
\addplot[black!40,mark=*,mark options={mark size=0.5pt,solid},only marks] table {images/modify-position/xy100};
\addplot[black!60,mark=*,mark options={mark size=0.5pt,solid},only marks] table {images/modify-position/xy200};
\addplot[black!80,mark=*,mark options={mark size=0.5pt,solid},only marks] table {images/modify-position/xy300};
\addplot[black,mark=*,mark options={mark size=0.5pt,solid},only marks] table {images/modify-position/xy400};
\addplot [black!40, dashed, domain=-65:-61.1, smooth] {0.00021659904003019882+4.8120506824188e-08*(x+63.55)};
\addplot [black!60, dashed, domain=-65:-61.1, smooth] {0.0003759371057439472+8.345020791175223e-08*(x+63.55)};
\addplot [black!80, dashed, domain=-65:-61.1, smooth] {0.0005006812108440317+1.1089442313278021e-07*(x+63.55)};
\addplot [black, dashed, domain=-65:-61.1, smooth] {0.0006007648732051406+1.327341470769046e-07*(x+63.55)};
\addlegendentry{100 iter.}
\addlegendentry{200 iter.}
\addlegendentry{300 iter.}
\addlegendentry{400 iter.}
\end{axis}
\end{tikzpicture}
\caption{Global perspective for various numbers of DROP steps. Changes in the set of traversed voxels are marked by vertical lines. The selected voxel is traversed by the selected track precisely if $x$ lies in the interval marked with thick vertical lines. At $\hat x=-63.55$, $f(x)$ is differentiable as verified in \cref{fig:ad-drop-track-coordinate-step-log} (in the case of 400~iterations), and the almost horizontal tangents are indicated by dashed lines.}
\label{fig:ad-drop-track-coordinate-global}
\end{subfigure} \\[0.3cm]
\begin{subfigure}[t]{0.4\textwidth}
\begin{tikzpicture}
\begin{axis}[xlabel={$x$},ylabel style={align=center},ylabel={$f(x)$},height=5cm,width=0.9\textwidth,%
xmin=-63.575,xmax=-61.575,
y tick label style={/pgf/number format/.cd,fixed,fixed zerofill,precision=3,/tikz/.cd},]
\addplot[black,mark=*,mark options={mark size=0.5pt,solid},only marks] table {images/modify-position/xy400-single-step};
\end{axis}
\end{tikzpicture}
\caption{Zoom into on of the steps of \cref{fig:ad-drop-track-coordinate-global}, with 400 iterations.}
\label{fig:ad-drop-track-coordinate-step}
\end{subfigure}
\qquad
\begin{subfigure}[t]{0.4\textwidth}
\begin{tikzpicture}
\begin{axis}[xlabel={$x-\hat x$},ylabel style={align=center},ylabel={$|f(x)-f(\hat x)|$},height=5cm,width=0.9\textwidth,%
ymode=log,xmode=log,]
\addplot[black,mark=*,mark options={mark size=0.5pt,solid},only marks] table {images/modify-position-log/xy400-diff};
\end{axis}
\end{tikzpicture}
\caption{Logarithmic plot of the numerator and denominator of the difference quotient in the same range as \cref{fig:ad-drop-track-coordinate-step}.}
\label{fig:ad-drop-track-coordinate-step-log}
\end{subfigure}
\caption{Dependency of a particular voxel's RSP on a particular track's beam spot coordinate. Elements of the system matrix $A$ were calculated using the mean chord length.}
\label{fig:ad-drop-track-coordinate}
\end{figure}
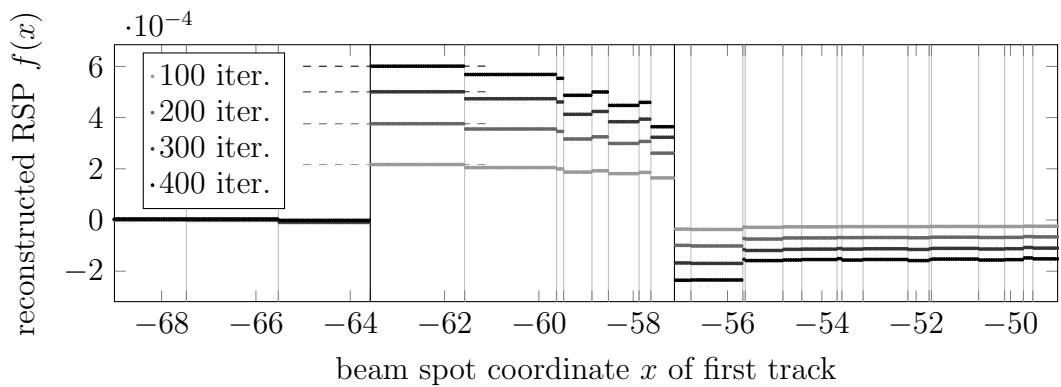
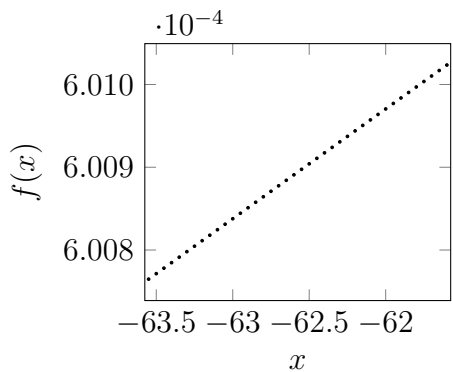
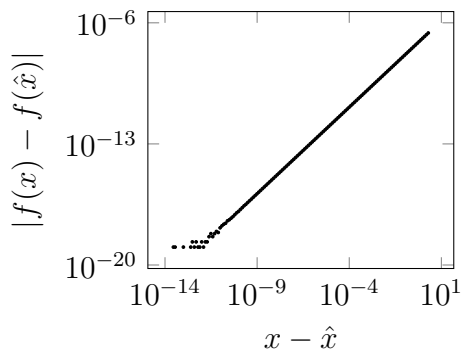

\section{Discussion: Challenges for Differentiating the Pipeline}\label{sec:pipeline-again}
As detailed in the end of \cref{sec:diff-algorithmic} based on \cref{fig:badapproxderiv}, algorithms designed without an AD option in mind might need manual adaptations to make sure that their derivatives approximate the ``true'' function's derivatives. Besides, the linearization~\eqref{eq:taylor} is only helpful for the quantification of uncertainties and for optimization if the true function is ``sufficiently smooth'', as discussed in \cref{sec:purpose-uq,sec:purpose-opt}.

In this section, we discuss these aspects for the software pipeline outlined in \cref{sec:software-pipeline}.

\subsection{Differentiation of Randomized Code}
No meaningful information could be gained by differentiating GATE with respect to its random input, i.\,e.\ the seed of the RNG. Regarding the other partial derivatives (w.\,r.\,t.\ detector parameters etc.), the RNG seed was kept constant in this study.

In \cref{sec:results-mc} we found that four particular outputs of GATE were piecewise differentiable with respect to the beam energy, but did also involve jumps. Two of the jumps were further investigated and it was found that the control flow of the program changed at this point because a floating-point comparison flipped, changing the number of calls to the RNG. Such a change severely affects the subsequent computations  because the program then receives a shifted sequence of random numbers. 

We therefore conjecture that the observed discontinuities are, at least partially, an artifact of how the RNG is used, and not of any physical significance. Restarting the RNG with precomputed random seeds at strategic locations in the code might remove some of this ``numerical chaos''. 

We should recall here that frequent discontinuities do not deteriorate the accuracy of AD for computing derivatives if the derivatives exist for the respective input. However, they diminish the accuracy of the linearization formula~\eqref{eq:taylor}, and therefore the value of the derivatives for applications in \cref{sec:purpose-uq,sec:purpose-opt} regarding stand-alone GATE. 
For the pipeline as a whole, the issue might be less pronounced, as subsequent computational steps combine simulations of many independent protons, possibly averaging out the chaotic behaviour while keeping systematic dependencies.
Further research in this direction may perturb inputs of a GATE simulation as part of a complete pCT software pipeline with a higher number of protons, and observe the reaction of, e.\,g., the reconstruction error of the RSP image.

Instead of applying AD to the particle physics simulator itself, simulation data can be used to fit a surrogate model, which is then differentiated instead of the simulator \cite{dorigo_toward_2022}. In general, we expect such an approach to reduce chaotic behaviour, evaluate faster, and reduce the workload needed to apply the AD tool, but it can be less accurate. Surrogate models for calorimeter showers are a very active field of research, see e.\,g.\ \citet{dorigo_toward_2022}.

\subsection{Discrete Variables Related to Detector Output}\label{sec:discrvar-detectoroutput}
The detector output consists of pixel activations that are either $0$ or $1$, i.\,e.\ take a \emph{discrete} value, as opposed to \emph{continuous} coordinates, energies etc. The continuous output of GATE is mapped into discrete values by the charge diffusion model's choice of which pixels to activate. 
The calculation of cluster centers preceding the track reconstructions maps the discrete pixel data back into a  continuous range. 

As the local behaviour of any function into a discrete set is either ``constant'' (not interesting) or ``having a step'' (not differentiable), we cannot make any use of derivative information here. Expressed differently, any discrete intermediate result comes from rounding of continuous coordinates, which erases all derivative information and is sometimes discontinuous. This is what we observed in \cref{sec:results-tracking}. \Cref{fig:tracking-diff-study-c} shows the jumps of \cref{fig:badapproxderiv-b}, and \cref{fig:tracking-diff-study-b} shows the noise of \cref{fig:badapproxderiv-c}. 

One idea to fix this is to replace the continuous-to-discrete-to-continuous conversion by a surrogate model. In the easiest case, one might just carry over the ``ground truth'' hit positions and energy depositions from the Monte Carlo subprocedure to the RSP reconstruction, bypassing the charge diffusion model, clustering and tracking subprocedures. 

\subsection{Numerical Noise of the MBIR Solver}
Even if the error of a least-squares solution provided by an approximative numerical solver is small, a noisy error as illustrated in \cref{fig:badapproxderiv-c}, and observed for LSCG in \cref{fig:ad-modify-wepl-lscg} in \cref{sec:results-mbir}, has large derivatives. The noise probably results from stopping the iterative solver before it reaches full convergence. DROP performs only linear operations on the right hand side and hence the reconstructed RSP depends on the WEPL in a linear way, making it a better choice for further investigations.

\subsection{Discrete Variables Related to the Matrix Generation in MBIR Algorithms}
While stepping along a path and determining the current voxel, an affine-linear function is applied to the current coordinates and the result is rounded; in the end, a certain path either intersects, or does not intersect, a certain voxel. This discrete choice introduces discontinuities w.\,r.\,t.\ track coordinates, as can be seen in \cref{fig:ad-drop-track-coordinate-global} in \cref{sec:results-mbir}, where the matrix element of an intersected voxel was set to a mean chord length. If a perturbation of a track coordinate does not change the set of voxels intersected by the path, this value only changes very little, due to its dependency on the tangent vector of the path. Therefore the reconstructed RSP is nearly a ``step function'' whose gradients exist by \cref{fig:ad-drop-track-coordinate-step,fig:ad-drop-track-coordinate-step-log}, but are useless for optimization and UQ.

In \cref{fig:fuzzy-ad-drop-track-coordinate} a fuzzy voxels approach was used, which leaves some discontinuities but makes the gradients represent the overall behaviour very well. This however comes at the price of a longer reconstruction time and blurrier result. 

Either way, output variables of the full pipeline might be smoother because they combine the RSP values of many voxels.

\section{Conclusions}\label{sec:conclusions}
We presented the algorithmic substeps of the \anon{Bergen pCT}{XXX} collaboration's incipient software pipeline, with special focus on linearizability as a prerequisite for gradient-based optimization and UQ.

Both the Monte Carlo and MBIR subprocedure's central steps compute piecewise differentiable functions with discontinuities. For the MBIR subprocedure, we identified the cause of discontinuities and proposed a way to mitigate it. Our study indicates that both codes are ready for the integration of AD. 

The proton history reconstruction subprocedure involves many discrete variables, which present a huge obstacle to (algorithmic) differentiability. We investigated the tracking step as an example and found a very noisy behaviour. It is probably the best approach to ``bridge'' this subprocedure, carrying over the ground truth.

\section{Acknowledgements}
\anon{We gratefully acknowledge the funding of the research training group SIVERT by the German federal state of Rhineland-Palatinate. 

This work is supported by the Research Council of Norway and the University of Bergen, grant number 250858; the Trond Mohn Foundation, grant number BFS2017TMT07; as well as the Hungarian NKFIH OTKA K135515 grant and the Wigner Scientific Computing Laboratory (WSCLAB).}{(removed for double-anonymous peer review)}

\section{Conflict of Interest Statement}
The authors have no relevant conflicts of interest to disclose.

\newcommand\newblock{}
\bibliography{pmb}

@article{jan_gate_2004,
	title = {{GATE} - {Geant4} {Application} for {Tomographic} {Emission}: a simulation toolkit for {PET} and {SPECT}},
	volume = {49},
	issn = {0031-9155},
	shorttitle = {{GATE} - {Geant4} {Application} for {Tomographic} {Emission}},
	url = {https://www.ncbi.nlm.nih.gov/pmc/articles/PMC3267383/},
	number = {19},
	urldate = {2020-10-19},
	journal = {Physics in Medicine and Biology},
	author = {Jan, S. and others},
	longauthor = {Jan, S. and Santin, G. and Strul, D. and Staelens, S. and Assié, K. and Autret, D. and Avner, S. and Barbier, R. and Bardiès, M. and Bloomfield, P. M. and Brasse, D. and Breton, V. and Bruyndonckx, P. and Buvat, I. and Chatziioannou, A. F. and Choi, Y. and Chung, Y. H. and Comtat, C. and Donnarieix, D. and Ferrer, L. and Glick, S. J. and Groiselle, C. J. and Guez, D. and Honore, P.-F. and Kerhoas-Cavata, S. and Kirov, A. S. and Kohli, V. and Koole, M. and Krieguer, M. and van der Laan, D. J. and Lamare, F. and Largeron, G. and Lartizien, C. and Lazaro, D. and Maas, M. C. and Maigne, L. and Mayet, F. and Melot, F. and Merheb, C. and Pennacchio, E. and Perez, J. and Pietrzyk, U. and Rannou, F. R. and Rey, M. and Schaart, D. R. and Schmidtlein, C. R. and Simon, L. and Song, T. Y. and Vieira, J.-M. and Visvikis, D. and Van de Walle, R. and Wieërs, E. and Morel, C.},
	month = oct,
	year = {2004},
	pmid = {15552416},
	pmcid = {PMC3267383},
	pages = {4543--4561},
}

@article{wohlfahrt_2020, title={Status and innovations in pre-treatment {CT} imaging for proton therapy}, volume={93}, ISSN={0007-1285, 1748-880X}, DOI={10.1259/bjr.20190590}, note={00004}, number={1107}, journal={The British Journal of Radiology}, author={Wohlfahrt, Patrick and Richter, Christian}, year={2020}, month={3}, pages={20190590} }

@article{allison_recent_2016,
	title = {Recent developments in {Geant4}},
	volume = {835},
	issn = {01689002},
	url = {https://linkinghub.elsevier.com/retrieve/pii/S0168900216306957},
	doi = {10.1016/j.nima.2016.06.125},
	language = {en},
	urldate = {2021-12-08},
	journal = {Nuclear Instruments and Methods in Physics Research Section A: Accelerators, Spectrometers, Detectors and Associated Equipment},
	author = {Allison, J. and others},
	longauthor = {Allison, J. and Amako, K. and Apostolakis, J. and Arce, P. and Asai, M. and Aso, T. and Bagli, E. and Bagulya, A. and Banerjee, S. and Barrand, G. and Beck, B.R. and Bogdanov, A.G. and Brandt, D. and Brown, J.M.C. and Burkhardt, H. and Canal, Ph. and Cano-Ott, D. and Chauvie, S. and Cho, K. and Cirrone, G.A.P. and Cooperman, G. and Cortés-Giraldo, M.A. and Cosmo, G. and Cuttone, G. and Depaola, G. and Desorgher, L. and Dong, X. and Dotti, A. and Elvira, V.D. and Folger, G. and Francis, Z. and Galoyan, A. and Garnier, L. and Gayer, M. and Genser, K.L. and Grichine, V.M. and Guatelli, S. and Guèye, P. and Gumplinger, P. and Howard, A.S. and Hřivnáčová, I. and Hwang, S. and Incerti, S. and Ivanchenko, A. and Ivanchenko, V.N. and Jones, F.W. and Jun, S.Y. and Kaitaniemi, P. and Karakatsanis, N. and Karamitros, M. and Kelsey, M. and Kimura, A. and Koi, T. and Kurashige, H. and Lechner, A. and Lee, S.B. and Longo, F. and Maire, M. and Mancusi, D. and Mantero, A. and Mendoza, E. and Morgan, B. and Murakami, K. and Nikitina, T. and Pandola, L. and Paprocki, P. and Perl, J. and Petrović, I. and Pia, M.G. and Pokorski, W. and Quesada, J.M. and Raine, M. and Reis, M.A. and Ribon, A. and Ristić Fira, A. and Romano, F. and Russo, G. and Santin, G. and Sasaki, T. and Sawkey, D. and Shin, J.I. and Strakovsky, I.I. and Taborda, A. and Tanaka, S. and Tomé, B. and Toshito, T. and Tran, H.N. and Truscott, P.R. and Urban, L. and Uzhinsky, V. and Verbeke, J.M. and Verderi, M. and Wendt, B.L. and Wenzel, H. and Wright, D.H. and Wright, D.M. and Yamashita, T. and Yarba, J. and Yoshida, H.},
	month = nov,
	year = {2016},
	pages = {186--225},
}

@article{allison_geant4_2006,
	title = {Geant4 developments and applications},
	volume = {53},
	issn = {0018-9499},
	url = {http://ieeexplore.ieee.org/document/1610988/},
	doi = {10.1109/TNS.2006.869826},
	number = {1},
	urldate = {2021-12-08},
	journal = {IEEE Transactions on Nuclear Science},
	author = {Allison, J. and others},
	longauthor = {Allison, J. and Amako, K. and Apostolakis, J. and Araujo, H. and Arce Dubois, P. and Asai, M. and Barrand, G. and Capra, R. and Chauvie, S. and Chytracek, R. and Cirrone, G.A.P. and Cooperman, G. and Cosmo, G. and Cuttone, G. and Daquino, G.G. and Donszelmann, M. and Dressel, M. and Folger, G. and Foppiano, F. and Generowicz, J. and Grichine, V. and Guatelli, S. and Gumplinger, P. and Heikkinen, A. and Hrivnacova, I. and Howard, A. and Incerti, S. and Ivanchenko, V. and Johnson, T. and Jones, F. and Koi, T. and Kokoulin, R. and Kossov, M. and Kurashige, H. and Lara, V. and Larsson, S. and Lei, F. and Link, O. and Longo, F. and Maire, M. and Mantero, A. and Mascialino, B. and McLaren, I. and Mendez Lorenzo, P. and Minamimoto, K. and Murakami, K. and Nieminen, P. and Pandola, L. and Parlati, S. and Peralta, L. and Perl, J. and Pfeiffer, A. and Pia, M.G. and Ribon, A. and Rodrigues, P. and Russo, G. and Sadilov, S. and Santin, G. and Sasaki, T. and Smith, D. and Starkov, N. and Tanaka, S. and Tcherniaev, E. and Tome, B. and Trindade, A. and Truscott, P. and Urban, L. and Verderi, M. and Walkden, A. and Wellisch, J.P. and Williams, D.C. and Wright, D. and Yoshida, H.},
	month = feb,
	year = {2006},
	pages = {270--278},
}

@article{agostinelli_geant4simulation_2003,
	title = {Geant4—a simulation toolkit},
	volume = {506},
	issn = {0168-9002},
	url = {http://www.sciencedirect.com/science/article/pii/S0168900203013688},
	doi = {10.1016/S0168-9002(03)01368-8},
	language = {en},
	number = {3},
	urldate = {2020-11-01},
	journal = {Nuclear Instruments and Methods in Physics Research Section A: Accelerators, Spectrometers, Detectors and Associated Equipment},
	author = {Agostinelli, S. and others},
	longauthor = {Agostinelli, S. and Allison, J. and Amako, K. and Apostolakis, J. and Araujo, H. and Arce, P. and Asai, M. and Axen, D. and Banerjee, S. and Barrand, G. and Behner, F. and Bellagamba, L. and Boudreau, J. and Broglia, L. and Brunengo, A. and Burkhardt, H. and Chauvie, S. and Chuma, J. and Chytracek, R. and Cooperman, G. and Cosmo, G. and Degtyarenko, P. and Dell'Acqua, A. and Depaola, G. and Dietrich, D. and Enami, R. and Feliciello, A. and Ferguson, C. and Fesefeldt, H. and Folger, G. and Foppiano, F. and Forti, A. and Garelli, S. and Giani, S. and Giannitrapani, R. and Gibin, D. and Gómez Cadenas, J. J. and González, I. and Gracia Abril, G. and Greeniaus, G. and Greiner, W. and Grichine, V. and Grossheim, A. and Guatelli, S. and Gumplinger, P. and Hamatsu, R. and Hashimoto, K. and Hasui, H. and Heikkinen, A. and Howard, A. and Ivanchenko, V. and Johnson, A. and Jones, F. W. and Kallenbach, J. and Kanaya, N. and Kawabata, M. and Kawabata, Y. and Kawaguti, M. and Kelner, S. and Kent, P. and Kimura, A. and Kodama, T. and Kokoulin, R. and Kossov, M. and Kurashige, H. and Lamanna, E. and Lampén, T. and Lara, V. and Lefebure, V. and Lei, F. and Liendl, M. and Lockman, W. and Longo, F. and Magni, S. and Maire, M. and Medernach, E. and Minamimoto, K. and Mora de Freitas, P. and Morita, Y. and Murakami, K. and Nagamatu, M. and Nartallo, R. and Nieminen, P. and Nishimura, T. and Ohtsubo, K. and Okamura, M. and O'Neale, S. and Oohata, Y. and Paech, K. and Perl, J. and Pfeiffer, A. and Pia, M. G. and Ranjard, F. and Rybin, A. and Sadilov, S. and Di Salvo, E. and Santin, G. and Sasaki, T. and Savvas, N. and Sawada, Y. and Scherer, S. and Sei, S. and Sirotenko, V. and Smith, D. and Starkov, N. and Stoecker, H. and Sulkimo, J. and Takahata, M. and Tanaka, S. and Tcherniaev, E. and Safai Tehrani, E. and Tropeano, M. and Truscott, P. and Uno, H. and Urban, L. and Urban, P. and Verderi, M. and Walkden, A. and Wander, W. and Weber, H. and Wellisch, J. P. and Wenaus, T. and Williams, D. C. and Wright, D. and Yamada, T. and Yoshida, H. and Zschiesche, D.},
	month = 7,
	year = {2003},
	pages = {250--303},
}

@article{tambave_2020, title={Characterization of monolithic CMOS pixel sensor chip with ion beams for application in particle computed tomography}, volume={958}, ISSN={01689002}, DOI={10.1016/j.nima.2019.162626}, abstractNote={Particle computed tomography (pCT) is an emerging imaging modality that promises to reduce range uncertainty in particle therapy. The Bergen pCT collaboration aims to develop a novel pCT prototype based on the ALPIDE monolithic CMOS sensor. The planned prototype consist of two tracking planes forming a rear tracker and Digital Tracking Calorimeter (DTC). The DTC will be made of a 41 layer ALPIDE-aluminum sandwich structure. To enable data acquisition at clinical particle rates, a large multiplicity of particles will be measured using the highly-granular ALPIDE sensor. In this work, a first characterization of the ALPIDE sensor performance in ion beams is conducted. Particle hits in the ALPIDE sensor result in charge clusters whose size is related to the chip response and the particle energy deposit. Firstly, measurements in a 10 MeV 4He micro beam have been conducted at the SIRIUS microprobe facility of ANSTO to investigate the dependence of the cluster size on the beam position over the ALPIDE pixel. Here, a variation in cluster size depending on the impinging point of the beam was observed. Additional beam tests were conducted at the Heidelberg Ion-Beam Therapy Center (HIT) investigating the cluster size as a function of the deposited energy by protons and 4He ions in the sensitive volume of the ALPIDE. Results show the expected increase in cluster sizes with deposited energy and a clear difference in cluster sizes for protons and 4He ions. As a conclusion, the variation in cluster size with the impinging point of the beam has to be accounted for to enable accurate energy loss reconstruction with the ALPIDE. This does, however, not affect the tracking of particles through the final prototype, as for that only the center-of-mass of the cluster is relevant.}, number={Proceedings of the Vienna Conference of Instrumentation 2019}, journal={Nuclear Instruments and Methods in Physics Research Section A: Accelerators, Spectrometers, Detectors and Associated Equipment}, author={Tambave, Ganesh and Alme, J. and Barnaföldi, G.G. and Barthel, R. and van den Brink, A. and Brons, S. and Chaar, M. and Eikeland, V. and Genov, G. and Grøttvik, O. and Pettersen, H.E.S. and Pastuovic, Z. and Huiberts, S. and Helstrup, H. and Hetland, K.F. and Mehendale, S. and Meric, I. and Malik, Q.W. and Odland, O.H. and Papp, G. and Peitzmann, T. and Piersimoni, P. and Ur Rehman, A. and Reidt, F. and Richter, M. and Röhrich, D. and Sudar, A. and Samnøy, A.T. and Seco, J. and Shafiee, H. and Skjæveland, E.V. and Sølie, J.R. and Ullaland, K. and Varga-Kofarago, M. and Volz, L. and Wagner, B. and Yang, S.}, year={2020}, month={4}, pages={162626} }

@article{pettersen_design_2019,
	title = {Design optimization of a pixel-based range telescope for proton computed tomography},
	volume = {63},
	issn = {11201797},
	url = {https://linkinghub.elsevier.com/retrieve/pii/S1120179719301358},
	doi = {10.1016/j.ejmp.2019.05.026},
	language = {en},
	urldate = {2021-12-08},
	journal = {Physica Medica},
	author = {Pettersen, Helge Egil Seime and others},
	longauthor = {Pettersen, Helge Egil Seime and Alme, Johan and Barnaföldi, Gergely Gábor and Barthel, Rene and van den Brink, Anthony and Chaar, Mamdouh and Eikeland, Viljar and García-Santos, Alba and Genov, Georgi and Grimstad, Silje and Grøttvik, Ola and Helstrup, Håvard and Hetland, Kristin Fanebust and Mehendale, Shruti and Meric, Ilker and Odland, Odd Harald and Papp, Gábor and Peitzmann, Thomas and Piersimoni, Pierluigi and Ur Rehman, Attiq and Richter, Matthias and Samnøy, Andreas Tefre and Seco, Joao and Shafiee, Hesam and Skjæveland, Eivind Vågslid and Sølie, Jarle Rambo and Tambave, Ganesh and Ullaland, Kjetil and Varga-Kofarago, Monika and Volz, Lennart and Wagner, Boris and Yang, Shiming and Röhrich, Dieter},
	month = 7,
	year = {2019},
	pages = {87--97},
}

@article{krah_comprehensive_2018,
	title = {A comprehensive theoretical comparison of proton imaging set-ups in terms of spatial resolution},
	volume = {63},
	issn = {1361-6560},
	url = {https://iopscience.iop.org/article/10.1088/1361-6560/aaca1f},
	doi = {10.1088/1361-6560/aaca1f},
	pages = {135013},
	number = {13},
	journaltitle = {Physics in Medicine \& Biology},
	journal = {Phys. Med. Biol.},
	author = {Krah, N and Khellaf, F and Létang, J M and Rit, S and Rinaldi, I},
	urldate = {2021-12-09},
	date = {2018-07-02},
  year={2018},
}

@article{schulte_maximum_2008,
	title = {A maximum likelihood proton path formalism for application in proton computed tomography: Maximum likelihood path formalism for proton {CT}},
	volume = {35},
	issn = {00942405},
	url = {http://doi.wiley.com/10.1118/1.2986139},
	doi = {10.1118/1.2986139},
	shorttitle = {A maximum likelihood proton path formalism for application in proton computed tomography},
	pages = {4849--4856},
	number = {11},
	journaltitle = {Medical Physics},
	journal = {Med. Phys.},
	author = {Schulte, R. W. and Penfold, S. N. and Tafas, J. T. and Schubert, K. E.},
	urldate = {2021-12-09},
	date = {2008-10-13},
	langid = {english},
  year={2008},
}

@article{lynch_approximations_1991,
	title = {Approximations to multiple Coulomb scattering},
	volume = {58},
	issn = {0168583X},
	url = {https://linkinghub.elsevier.com/retrieve/pii/0168583X9195671Y},
	doi = {10.1016/0168-583X(91)95671-Y},
	pages = {6--10},
	number = {1},
	journaltitle = {Nuclear Instruments and Methods in Physics Research Section B: Beam Interactions with Materials and Atoms},
	journal = {Nuclear Instruments and Methods in Physics Research Section B: Beam Interactions with Materials and Atoms},
	author = {Lynch, Gerald R. and Dahl, Orin I.},
	urldate = {2021-12-09},
	date = {1991-05},
	langid = {english},
  year={1991},
}

@article{gottschalk_multiple_1993,
	title = {Multiple Coulomb scattering of 160 {MeV} protons},
	volume = {74},
	issn = {0168583X},
	url = {https://linkinghub.elsevier.com/retrieve/pii/0168583X9395944Z},
	doi = {10.1016/0168-583X(93)95944-Z},
	pages = {467--490},
	number = {4},
	journaltitle = {Nuclear Instruments and Methods in Physics Research Section B: Beam Interactions with Materials and Atoms},
	journal = {Nuclear Instruments and Methods in Physics Research Section B: Beam Interactions with Materials and Atoms},
	author = {Gottschalk, B. and Koehler, A.M. and Schneider, R.J. and Sisterson, J.M. and Wagner, M.S.},
	urldate = {2021-12-09},
	date = {1993-06},
	langid = {english},
  year={1993},
}

@article{alme_high-granularity_2020,
	title = {A High-Granularity Digital Tracking Calorimeter Optimized for Proton {CT}},
	volume = {8},
	issn = {2296-424X},
	url = {https://www.frontiersin.org/article/10.3389/fphy.2020.568243},
	doi = {10.3389/fphy.2020.568243},
	pages = {460},
	journaltitle = {Frontiers in Physics},
	author = {Alme, Johan and others},
	longauthor = {Alme, Johan and Barnaföldi, Gergely Gábor and Barthel, Rene and Borshchov, Vyacheslav and Bodova, Tea and van den Brink, Anthony and Brons, Stephan and Chaar, Mamdouh and Eikeland, Viljar and Feofilov, Grigory and Genov, Georgi and Grimstad, Silje and Grøttvik, Ola and Helstrup, Håvard and Herland, Alf and Hilde, Annar Eivindplass and Igolkin, Sergey and Keidel, Ralf and Kobdaj, Chinorat and van der Kolk, Naomi and Listratenko, Oleksandr and Malik, Qasim Waheed and Mehendale, Shruti and Meric, Ilker and Nesbø, Simon Voigt and Odland, Odd Harald and Papp, Gábor and Peitzmann, Thomas and Seime Pettersen, Helge Egil and Piersimoni, Pierluigi and Protsenko, Maksym and Rehman, Attiq Ur and Richter, Matthias and Röhrich, Dieter and Samnøy, Andreas Tefre and Seco, Joao and Setterdahl, Lena and Shafiee, Hesam and Skjolddal, Øistein Jelmert and Solheim, Emilie and Songmoolnak, Arnon and Sudár, Ákos and Sølie, Jarle Rambo and Tambave, Ganesh and Tymchuk, Ihor and Ullaland, Kjetil and Underdal, Håkon Andreas and Varga-Köfaragó, Monika and Volz, Lennart and Wagner, Boris and Widerøe, Fredrik Mekki and Xiao, {RenZheng} and Yang, Shiming and Yokoyama, Hiroki},
	date = {2020},
}

@article{pettersen_proton_2020,
	title = {Proton Tracking Algorithm in a Pixel-Based Range Telescope for Proton Computed Tomography},
	url = {http://arxiv.org/abs/2006.09751},
	journal = {{arXiv}:2006.09751 [physics]},
	author = {Pettersen, Helge Egil Seime and Meric, Ilker and Odland, Odd Harald and Shafiee, Hesam and Sølie, Jarle Rambo and Röhrich, Dieter},
	urldate = {2020-12-03},
	date = {2020-06-17},
	eprinttype = {arxiv},
	eprint = {2006.09751},
  year={2020},
}

@article{strandlie_track_2010,
	title = {Track and vertex reconstruction: From classical to adaptive methods},
	volume = {82},
	issn = {0034-6861, 1539-0756},
	url = {https://link.aps.org/doi/10.1103/RevModPhys.82.1419},
	doi = {10.1103/RevModPhys.82.1419},
	shorttitle = {Track and vertex reconstruction},
	pages = {1419--1458},
	number = {2},
	journaltitle = {Reviews of Modern Physics},
	journal = {Rev. Mod. Phys.},
	author = {Strandlie, Are and Frühwirth, Rudolf},
	urldate = {2020-07-03},
	date = {2010-05-07},
	langid = {english},
  year={2010},
}

@article{bortfeld_analytical_1997,
	title = {An analytical approximation of the Bragg curve for therapeutic proton beams},
	volume = {24},
	issn = {00942405},
	url = {http://doi.wiley.com/10.1118/1.598116},
	doi = {10.1118/1.598116},
	pages = {2024--2033},
	number = {12},
	journaltitle = {Medical Physics},
	journal = {Med. Phys.},
	author = {Bortfeld, Thomas},
	urldate = {2021-12-10},
	date = {1997-12},
	langid = {english},
  year={1997},
}

@article{pettersen_proton_2017,
	title = {Proton tracking in a high-granularity Digital Tracking Calorimeter for proton {CT} purposes},
	volume = {860},
	issn = {01689002},
	url = {https://linkinghub.elsevier.com/retrieve/pii/S0168900217301882},
	doi = {10.1016/j.nima.2017.02.007},
	pages = {51--61},
	journaltitle = {Nuclear Instruments and Methods in Physics Research Section A: Accelerators, Spectrometers, Detectors and Associated Equipment},
	journal = {Nuclear Instruments and Methods in Physics Research Section A: Accelerators, Spectrometers, Detectors and Associated Equipment},
	author = {Pettersen, Helge Egil Seime and others},
	longauthor = {Pettersen, Helge Egil Seime and Alme, J. and Biegun, A. and van den Brink, A. and Chaar, M. and Fehlker, D. and Meric, I. and Odland, O.H. and Peitzmann, T. and Rocco, E. and Ullaland, K. and Wang, H. and Yang, S. and Zhang, C. and Röhrich, D.},
	urldate = {2019-05-21},
	date = {2017-07},
	langid = {english},
  year={2017},
}

@article{pettersen_investigating_2021, 
   title={Investigating particle track topology for range telescopes in particle radiography using convolutional neural networks}, volume={60}, ISSN={0284-186X, 1651-226X}, DOI={10.1080/0284186X.2021.1949037}, number={11}, journal={Acta Oncologica}, 
   author={Pettersen, Helge Egil Seime and others},
   longauthor={Pettersen, Helge Egil Seime and Aehle, Max and Alme, Johan and Barnaföldi, Gergely Gábor and Borshchov, Vyacheslav and van den Brink, Anthony and Chaar, Mamdouh and Eikeland, Viljar and Feofilov, Grigory and Garth, Christoph and Gauger, Nicolas R. and Genov, Georgi and Grøttvik, Ola and Helstrup, Håvard and Igolkin, Sergey and Keidel, Ralf and Kobdaj, Chinorat and Kortus, Tobias and Leonhardt, Viktor and Mehendale, Shruti and Mulawade, Raju Ningappa and Odland, Odd Harald and Papp, Gábor and Peitzmann, Thomas and Piersimoni, Pierluigi and Protsenko, Maksym and Rehman, Attiq Ur and Richter, Matthias and Santana, Joshua and Schilling, Alexander and Seco, Joao and Songmoolnak, Arnon and Sølie, Jarle Rambo and Tambave, Ganesh and Tymchuk, Ihor and Ullaland, Kjetil and Varga-Kofarago, Monika and Volz, Lennart and Wagner, Boris and Wendzel, Steffen and Wiebel, Alexander and Xiao, RenZheng and Yang, Shiming and Yokoyama, Hiroki and Zillien, Sebastian and Röhrich, Dieter}, 
   year={2021}, month={7}, pages={1413–1418} }

@article{pettersen_accuracy_2018, title={Accuracy of parameterized proton range models; a comparison}, volume={144}, DOI={10.1016/j.radphyschem.2017.08.028}, abstractNote={An accurate calculation of proton ranges in phantoms or detector geometries is crucial for decision making in proton
therapy and proton imaging. To this end, several parameterizations of the range-energy relationship exist, with
different levels of complexity and accuracy. In this study we compare the accuracy four different parameterizations
models: Two analytical models derived from the Bethe equation, and two different interpolation schemes applied to range-energy tables. In conclusion, a spline interpolation scheme yields the highest reproduction accuracy, while the
shape of the energy loss-curve is best reproduced with the differentiated Bragg-Kleeman equation.}, note={IF 1.984}, journal={Radiation Physics and Chemistry}, author={Pettersen, Helge Egil Seime and Chaar, M. and Meric, I. and Odland, O. H. and Sølie, J. R. and Röhrich, D.}, year={2018}, month={3}, pages={295–297} }

@article{Collins-Fekete_theoretical_2017, title={A theoretical framework to predict the most likely ion path in particle imaging}, volume={62}, ISSN={0031-9155, 1361-6560}, DOI={10.1088/1361-6560/aa58ce}, number={5}, journal={Physics in Medicine \& Biology}, author={Collins-Fekete, Charles-Antoine and Volz, Lennart and Portillo, Stephen K N and Beaulieu, Luc and Seco, Joao}, year={2017}, month={3}, pages={1777–1790} }

@unpublished{aehle_quantification_2021,
 author = {Aehle, Max and Leonhardt, Viktor},
	title = {Quantification and Visualization of Uncertainties in {CT} Reconstruction},
	url = {http://www.ionimaging.org/llu2021-overview/},
	note = {7th Annual Loma Linda Workshop},
	urldate = {2021-12-10},
	date = {2021-08-03},
  year={2021}
}

@article{rios_derivative-free_2013,
	title = {Derivative-free optimization: a review of algorithms and comparison of software implementations},
	volume = {56},
	issn = {0925-5001, 1573-2916},
	url = {https://link.springer.com/10.1007/s10898-012-9951-y},
	doi = {10.1007/s10898-012-9951-y},
	shorttitle = {Derivative-free optimization},
	pages = {1247--1293},
	number = {3},
	journaltitle = {Journal of Global Optimization},
	journal = {J Glob Optim},
	author = {Rios, Luis Miguel and Sahinidis, Nikolaos V.},
	urldate = {2021-12-10},
	date = {2013-07},
	langid = {english},
  year={2013},
}

@article{larson_derivative-free_2019,
	title = {Derivative-free optimization methods},
	volume = {28},
	issn = {0962-4929, 1474-0508},
	url = {http://arxiv.org/abs/1904.11585},
	doi = {10.1017/S0962492919000060},
	pages = {287--404},
	journal = {Acta Numerica},
	author = {Larson, Jeffrey and Menickelly, Matt and Wild, Stefan M.},
	urldate = {2021-12-10},
	date = {2019-05-01},
	eprinttype = {arxiv},
	eprint = {1904.11585},
  year={2019},
}

@article{baydin_toward_2021,
	title = {Toward Machine Learning Optimization of Experimental Design},
	volume = {31},
	issn = {1061-9127, 1931-7336},
	url = {https://www.tandfonline.com/doi/full/10.1080/10619127.2021.1881364},
	doi = {10.1080/10619127.2021.1881364},
	pages = {25--28},
	number = {1},
	journaltitle = {Nuclear Physics News},
	journal = {Nuclear Physics News},
	author = {Baydin, Atılım Güneş and others},
	longauthor = {Baydin, Atılım Güneş and Cranmer, Kyle and Manzano, Pablo de Castro and Delaere, Christophe and Derkach, Denis and Donini, Julien and Dorigo, Tommaso and Giammanco, Andrea and Kieseler, Jan and Layer, Lukas and Louppe, Gilles and Ratnikov, Fedor and Strong, Giles C. and Tosi, Mia and Ustyuzhanin, Andrey and Vischia, Pietro and Yarar, Hevjin},
	urldate = {2021-12-11},
	date = {2021-01-02},
	langid = {english},
  year={2021},
}

@article{Yang_2012,
	doi = {10.1088/0031-9155/57/13/4095},
	url = {https://doi.org/10.1088/0031-9155/57/13/4095},
	year = 2012,
	month = {6},
	publisher = {{IOP} Publishing},
	volume = {57},
	number = {13},
	pages = {4095--4115},
	author = {Ming Yang and X Ronald Zhu and Peter C Park and Uwe Titt and Radhe Mohan and Gary Virshup and James E Clayton and Lei Dong},
	title = {Comprehensive analysis of proton range uncertainties related to patient stopping-power-ratio estimation using the stoichiometric calibration},
	journal = {Physics in Medicine and Biology},
}

@article{Paganetti_2012,
	doi = {10.1088/0031-9155/57/11/r99},
	url = {https://doi.org/10.1088/0031-9155/57/11/r99},
	year = 2012,
	month = {5},
	publisher = {{IOP} Publishing},
	volume = {57},
	number = {11},
	pages = {R99--R117},
	author = {Harald Paganetti},
	title = {Range uncertainties in proton therapy and the role of Monte Carlo simulations},
	journal = {Physics in Medicine and Biology},
}

@article{yang_theoretical_2010,
	title = {Theoretical variance analysis of single- and dual-energy computed tomography methods for calculating proton stopping power ratios of biological tissues},
	volume = {55},
	issn = {0031-9155, 1361-6560},
	url = {https://iopscience.iop.org/article/10.1088/0031-9155/55/5/006},
	doi = {10.1088/0031-9155/55/5/006},
	pages = {1343--1362},
	number = {5},
	journaltitle = {Physics in Medicine and Biology},
	journal = {Phys. Med. Biol.},
	author = {Yang, M and Virshup, G and Clayton, J and Zhu, X R and Mohan, R and Dong, L},
	urldate = {2021-12-11},
	date = {2010-03-07},
  year={2010},
}

@article{dedes_experimental_2019,
	title = {Experimental comparison of proton {CT} and dual energy x-ray {CT} for relative stopping power estimation in proton therapy},
	volume = {64},
	issn = {1361-6560},
	url = {https://iopscience.iop.org/article/10.1088/1361-6560/ab2b72},
	doi = {10.1088/1361-6560/ab2b72},
	pages = {165002},
	number = {16},
	journaltitle = {Physics in Medicine \& Biology},
	journal = {Phys. Med. Biol.},
	author = {Dedes, George and others},
	longauthor = {Dedes, George and Dickmann, Jannis and Niepel, Katharina and Wesp, Philipp and Johnson, Robert P and Pankuch, Mark and Bashkirov, Vladimir and Rit, Simon and Volz, Lennart and Schulte, Reinhard W and Landry, Guillaume and Parodi, Katia},
	urldate = {2021-12-11},
	date = {2019-08-14},
  year={2019}
}

@INCOLLECTION{Walther2012Gsw,
       title = "Getting started with {ADOL-C}",
       author = "A. Walther and A. Griewank",
       editor = "U. Naumann and O. Schenk",
       publisher = "Chapman-Hall CRC Computational Science",
       year = "2012",
       booktitle = "Combinatorial Scientific Computing",
       pages = "181--202",
       chapter = "7",
       ad_tools = "ADOL-C"
}

@article{SaAlGauTOMS2019,
title = {High-Performance Derivative Computations using CoDiPack},
author = {M. Sagebaum and T. Albring and N.R. Gauger},
url = {https://dl.acm.org/doi/abs/10.1145/3356900},
year = {2019},
date = {2019-12-01},
journal = {ACM Transactions on Mathematical Software (TOMS)},
volume = {45},
number = {4},
pubstate = {published},
tppubtype = {article},
}

@article{naumann_adjoint_2018,
	location = {Rochester, {NY}},
	title = {Adjoint Algorithmic Differentiation Tool Support for Typical Numerical Patterns in Computational Finance},
	url = {https://papers.ssrn.com/abstract=3122293},
  volume = {21},
	number = {4},
	institution = {Social Science Research Network},
	journal = {{SSRN} Scholarly Paper},
	author = {Naumann, Uwe and Toit, Jack},
	urldate = {2021-12-13},
	date = {2018-02-12},
	langid = {english},
  year={2018},
}

@article{solie_imagequality_2020, title={Image quality of list-mode proton imaging without front trackers}, volume={65}, ISSN={1361-6560}, DOI={10.1088/1361-6560/ab8ddb}, note={00000}, number={13}, journal={Physics in Medicine \& Biology}, 
  author={Sølie, Jarle Rambo and others},
  longauthor={Sølie, Jarle Rambo and Volz, Lennart and Pettersen, Helge Egil Seime and Piersimoni, Pierluigi and Odland, Odd Harald and Röhrich, Dieter and Helstrup, Håvard and Peitzmann, Thomas and Ullaland, Kjetil and Varga-Kofarago, Monika and Mehendale, Shruti and Grøttvik, Ola Slettevoll and Eikeland, Viljar Nilsen and Meric, Ilker and Seco, Joao}, 
  year={2020}, month={7}, pages={135012} }

@article{penfold_total_2010,
	title = {Total variation superiorization schemes in proton computed tomography image reconstruction: Total variation superiorization in proton {CT}},
	volume = {37},
	issn = {00942405},
	url = {http://doi.wiley.com/10.1118/1.3504603},
	doi = {10.1118/1.3504603},
	shorttitle = {Total variation superiorization schemes in proton computed tomography image reconstruction},
	pages = {5887--5895},
	number = {11},
	journaltitle = {Medical Physics},
	journal = {Med. Phys.},
	author = {Penfold, S. N. and Schulte, R. W. and Censor, Y. and Rosenfeld, A. B.},
	urldate = {2021-12-15},
	date = {2010-10-20},
	langid = {english},
  year={2010},
}

@article{biguri_tigre_2016,
	title = {{TIGRE}: a {MATLAB}-{GPU} toolbox for {CBCT} image reconstruction},
	volume = {2},
	issn = {2057-1976},
	url = {https://iopscience.iop.org/article/10.1088/2057-1976/2/5/055010},
	doi = {10.1088/2057-1976/2/5/055010},
	shorttitle = {{TIGRE}},
	pages = {055010},
	number = {5},
	journaltitle = {Biomedical Physics \& Engineering Express},
	journal = {Biomed. Phys. Eng. Express},
	author = {Biguri, Ander and Dosanjh, Manjit and Hancock, Steven and Soleimani, Manuchehr},
	urldate = {2021-12-15},
	date = {2016-09-08},
  year={2016},
}

@article{van_aarle_astra_2015,
	title = {The {ASTRA} Toolbox: A platform for advanced algorithm development in electron tomography},
	volume = {157},
	issn = {03043991},
	url = {https://linkinghub.elsevier.com/retrieve/pii/S0304399115001060},
	doi = {10.1016/j.ultramic.2015.05.002},
	shorttitle = {The {ASTRA} Toolbox},
	pages = {35--47},
	journaltitle = {Ultramicroscopy},
	journal = {Ultramicroscopy},
	author = {van Aarle, Wim and Palenstijn, Willem Jan and De Beenhouwer, Jan and Altantzis, Thomas and Bals, Sara and Batenburg, K. Joost and Sijbers, Jan},
	urldate = {2021-12-15},
	date = {2015-10},
	langid = {english},
  year={2015},
}

@unpublished{anderson_econ_nodate,
	location = {University of California, Berkeley},
	title = {Econ 204: Taylor’s Theorem},
	url = {https://eml.berkeley.edu/~anderson/Econ204/TaylorsTheoremTimeless.pdf},
	note = {Lecture Notes},
	author = {Anderson, Robert M.},
	urldate = {2021-12-15},
  year={2021},
}

@misc{tensorflow2015-whitepaper,
title={ {TensorFlow}: Large-Scale Machine Learning on Heterogeneous Systems},
url={https://www.tensorflow.org/},
note={Software available from tensorflow.org},
author={ Martín~Abadi and others},
longauthor={
    Martín~Abadi and
    Ashish~Agarwal and
    Paul~Barham and
    Eugene~Brevdo and
    Zhifeng~Chen and
    Craig~Citro and
    Greg~S.~Corrado and
    Andy~Davis and
    Jeffrey~Dean and
    Matthieu~Devin and
    Sanjay~Ghemawat and
    Ian~Goodfellow and
    Andrew~Harp and
    Geoffrey~Irving and
    Michael~Isard and
    Yangqing Jia and
    Rafal~Jozefowicz and
    Lukasz~Kaiser and
    Manjunath~Kudlur and
    Josh~Levenberg and
    Dandelion~Mané and
    Rajat~Monga and
    Sherry~Moore and
    Derek~Murray and
    Chris~Olah and
    Mike~Schuster and
    Jonathon~Shlens and
    Benoit~Steiner and
    Ilya~Sutskever and
    Kunal~Talwar and
    Paul~Tucker and
    Vincent~Vanhoucke and
    Vijay~Vasudevan and
    Fernanda~Viégas and
    Oriol~Vinyals and
    Pete~Warden and
    Martin~Wattenberg and
    Martin~Wicke and
    Yuan~Yu and
    Xiaoqiang~Zheng},
  year={2015},
}

@incollection{NEURIPS2019_9015,
title = {PyTorch: An Imperative Style, High-Performance Deep Learning Library},
author = {Paszke, Adam and others},
longauthor = {Paszke, Adam and Gross, Sam and Massa, Francisco and Lerer, Adam and Bradbury, James and Chanan, Gregory and Killeen, Trevor and Lin, Zeming and Gimelshein, Natalia and Antiga, Luca and Desmaison, Alban and Kopf, Andreas and Yang, Edward and DeVito, Zachary and Raison, Martin and Tejani, Alykhan and Chilamkurthy, Sasank and Steiner, Benoit and Fang, Lu and Bai, Junjie and Chintala, Soumith},
booktitle = {Advances in Neural Information Processing Systems 32},
editor = {H. Wallach and H. Larochelle and A. Beygelzimer and F. d'Alché-Buc and E. Fox and R. Garnett},
pages = {8024--8035},
year = {2019},
publisher = {Curran Associates, Inc.},
url = {http://papers.neurips.cc/paper/9015-pytorch-an-imperative-style-high-performance-deep-learning-library.pdf}
}

@article{penfold_more_2009,
	title = {A more accurate reconstruction system matrix for quantitative proton computed tomography: Reconstruction system matrix for quantitative proton {CT}},
	volume = {36},
	issn = {00942405},
	url = {http://doi.wiley.com/10.1118/1.3218759},
	doi = {10.1118/1.3218759},
	shorttitle = {A more accurate reconstruction system matrix for quantitative proton computed tomography},
	pages = {4511--4518},
	number = {10},
	journaltitle = {Medical Physics},
	journal = {Med. Phys.},
	author = {Penfold, S. N. and Rosenfeld, A. B. and Schulte, R. W. and Schubert, K. E.},
	urldate = {2021-12-15},
	date = {2009-09-09},
	langid = {english},
  year={2009},
}

@article{penfold_techniques_2015,
	title = {Techniques in Iterative Proton {CT} Image Reconstruction},
	volume = {16},
	issn = {1557-2064, 1557-2072},
	url = {http://link.springer.com/10.1007/s11220-015-0122-3},
	doi = {10.1007/s11220-015-0122-3},
	pages = {19},
	number = {1},
	journaltitle = {Sensing and Imaging},
	journal = {Sens Imaging},
	author = {Penfold, Scott and Censor, Yair},
	urldate = {2021-12-15},
	date = {2015-12},
	langid = {english},
  year={2015},
}

@book{griewank_evaluating_2008,
	title = {Evaluating Derivatives},
	isbn = {9780898716597},
	url = {https://epubs.siam.org/doi/book/10.1137/1.9780898717761},
	series = {Other Titles in Applied Mathematics},
	pagetotal = {448},
	publisher = {Society for Industrial and Applied Mathematics},
	author = {Griewank, Andreas and Walther, Andrea},
	urldate = {2022-01-06},
	date = {2008-01-01},
	year={2008},
	doi = {10.1137/1.9780898717761},
}

@book{naumann_art_2011,
	title = {The Art of Differentiating Computer Programs: An Introduction to Algorithmic Differentiation},
	isbn = {9781611972078},
	url = {http://epubs.siam.org/doi/book/10.1137/1.9781611972078},
	shorttitle = {The Art of Differentiating Computer Programs},
	publisher = {Society for Industrial and Applied Mathematics},
	author = {Naumann, Uwe},
	urldate = {2022-01-06},
	date = {2011-01},
	year={2011},
	langid = {english},
	doi = {10.1137/1.9781611972078},
}

@article{hurley_water-equivalent_2012,
	title = {Water-equivalent path length calibration of a prototype proton {CT} scanner: Water-equivalent path length calibration for proton {CT}},
	volume = {39},
	issn = {00942405},
	url = {http://doi.wiley.com/10.1118/1.3700173},
	doi = {10.1118/1.3700173},
	shorttitle = {Water-equivalent path length calibration of a prototype proton {CT} scanner},
	pages = {2438--2446},
	number = {5},
	journaltitle = {Medical Physics},
	journal = {Med. Phys.},
	author = {Hurley, R. F. and Schulte, R. W. and Bashkirov, V. A. and Wroe, A. J. and Ghebremedhin, A. and Sadrozinski, H. F.-W. and Rykalin, V. and Coutrakon, G. and Koss, P. and Patyal, B.},
	urldate = {2022-01-24},
	date = {2012-04-13},
	langid = {english},
  year={2012},
}

@article{meyer_optimization_2020,
	title = {Optimization and performance study of a proton {CT} system for pre-clinical small animal imaging},
	volume = {65},
	issn = {1361-6560},
	url = {https://iopscience.iop.org/article/10.1088/1361-6560/ab8afc},
	doi = {10.1088/1361-6560/ab8afc},
	pages = {155008},
	number = {15},
	journaltitle = {Physics in Medicine \& Biology},
	journal = {Phys. Med. Biol.},
	author = {Meyer, Sebastian and Bortfeldt, Jonathan and Lämmer, Paulina and Englbrecht, Franz S and Pinto, Marco and Schnürle, Katrin and Würl, Matthias and Parodi, Katia},
	urldate = {2022-01-21},
	date = {2020-08-13},
  year={2020},
}

@article{esposito_pravda_2018,
	title = {{PRaVDA}: The first solid-state system for proton computed tomography},
	volume = {55},
	issn = {11201797},
	url = {https://linkinghub.elsevier.com/retrieve/pii/S1120179718313073},
	doi = {10.1016/j.ejmp.2018.10.020},
	shorttitle = {{PRaVDA}},
	pages = {149--154},
	journaltitle = {Physica Medica},
	journal = {Physica Medica},
	author = {Esposito, Michela and others},
	longauthor = {Esposito, Michela and Waltham, Chris and Taylor, Jonathan T. and Manger, Sam and Phoenix, Ben and Price, Tony and Poludniowski, Gavin and Green, Stuart and Evans, Philip M. and Allport, Philip P. and Manolopulos, Spyros and Nieto-Camero, Jaime and Symons, Julyan and Allinson, Nigel M.},
	urldate = {2022-01-25},
	date = {2018-11},
	langid = {english},
  year={2018},
}

@article{scaringella_prima_2013,
	title = {The {PRIMA} ({PRoton} {IMAging}) collaboration: Development of a proton Computed Tomography apparatus},
	volume = {730},
	issn = {01689002},
	url = {https://linkinghub.elsevier.com/retrieve/pii/S0168900213008036},
	doi = {10.1016/j.nima.2013.05.181},
	shorttitle = {The {PRIMA} ({PRoton} {IMAging}) collaboration},
	pages = {178--183},
	journaltitle = {Nuclear Instruments and Methods in Physics Research Section A: Accelerators, Spectrometers, Detectors and Associated Equipment},
	journal = {Nuclear Instruments and Methods in Physics Research Section A: Accelerators, Spectrometers, Detectors and Associated Equipment},
	author = {Scaringella, M. and others},
	longauthor = {Scaringella, M. and Brianzi, M. and Bruzzi, M. and Bucciolini, M. and Carpinelli, M. and Cirrone, G.A.P. and Civinini, C. and Cuttone, G. and Lo Presti, D. and Pallotta, S. and Pugliatti, C. and Randazzo, N. and Romano, F. and Sipala, V. and Stancampiano, C. and Talamonti, C. and Tesi, M. and Vanzi, E. and Zani, M.},
	urldate = {2022-01-25},
	date = {2013-12},
	langid = {english},
  year={2013},
}

@article{saraya_study_2014,
	title = {Study of spatial resolution of proton computed tomography using a silicon strip detector},
	volume = {735},
	issn = {01689002},
	url = {https://linkinghub.elsevier.com/retrieve/pii/S0168900213012850},
	doi = {10.1016/j.nima.2013.09.051},
	pages = {485--489},
	journaltitle = {Nuclear Instruments and Methods in Physics Research Section A: Accelerators, Spectrometers, Detectors and Associated Equipment},
	journal = {Nuclear Instruments and Methods in Physics Research Section A: Accelerators, Spectrometers, Detectors and Associated Equipment},
	author = {Saraya, Y. and Izumikawa, T. and Goto, J. and Kawasaki, T. and Kimura, T.},
	urldate = {2022-01-25},
	date = {2014-01},
	langid = {english},
  year={2014},
}

@article{naimuddin_development_2016,
	title = {Development of a proton Computed Tomography detector system},
	volume = {11},
	issn = {1748-0221},
	url = {https://iopscience.iop.org/article/10.1088/1748-0221/11/02/C02012},
	doi = {10.1088/1748-0221/11/02/C02012},
	pages = {C02012--C02012},
	number = {2},
	journaltitle = {Journal of Instrumentation},
	journal = {J. Inst.},
	author = {Naimuddin, Md. and others},
	longauthor = {Naimuddin, Md. and Coutrakon, G. and Blazey, G. and Boi, S. and Dyshkant, A. and Erdelyi, B. and Hedin, D. and Johnson, E. and Krider, J. and Rukalin, V. and Uzunyan, S.A. and Zutshi, V. and Fordt, R. and Sellberg, G. and Rauch, J.E. and Roman, M. and Rubinov, P. and Wilson, P.},
	urldate = {2022-01-25},
	date = {2016-02-04},
  year={2016},
}

@article{sirkes_finite_1997,
	title = {Finite Difference of Adjoint or Adjoint of Finite Difference?},
	volume = {125},
	issn = {0027-0644, 1520-0493},
	url = {http://journals.ametsoc.org/doi/10.1175/1520-0493(1997)125<3373:FDOAOA>2.0.CO;2},
	doi = {10.1175/1520-0493(1997)125<3373:FDOAOA>2.0.CO;2},
	pages = {3373--3378},
	number = {12},
	journaltitle = {Monthly Weather Review},
	journal = {Mon. Wea. Rev.},
	author = {Sirkes, Ziv and Tziperman, Eli},
	urldate = {2022-01-25},
	date = {1997-12},
	langid = {english},
  year={1997},
}

@incollection{hutchison_data-flow_2006,
	location = {Berlin, Heidelberg},
	title = {The Data-Flow Equations of Checkpointing in Reverse Automatic Differentiation},
	volume = {3994},
	pages = {566--573},
	booktitle = {Computational Science – {ICCS} 2006},
	publisher = {Springer Berlin Heidelberg},
	author = {Dauvergne, Benjamin and Hascoët, Laurent},
	editor = {Alexandrov, Vassil N. and van Albada, Geert Dick and Sloot, Peter M. A. and Dongarra, Jack},
	editorb = {Hutchison, David and Kanade, Takeo and Kittler, Josef and Kleinberg, Jon M. and Mattern, Friedemann and Mitchell, John C. and Naor, Moni and Nierstrasz, Oscar and Pandu Rangan, C. and Steffen, Bernhard and Sudan, Madhu and Terzopoulos, Demetri and Tygar, Dough and Vardi, Moshe Y. and Weikum, Gerhard},
	editorbtype = {redactor},
	urldate = {2022-01-26},
	date = {2006},
	year={2006},
	doi = {10.1007/11758549_78},
}

@inproceedings{maclaurin2015autograd,
  title={Autograd: Effortless gradients in numpy},
  author={Maclaurin, Dougal and Duvenaud, David and Adams, Ryan P},
  booktitle={ICML 2015 AutoML Workshop},
  volume={238},
  pages={5},
  year={2015}
}

@article{Albring_etal2016b,
title = {Efficient Aerodynamic Design using the Discrete Adjoint Method in SU2},
author = {T. Albring and M. Sagebaum and N.R. Gauger},
year = {2016},
date = {2016-07-02},
journal = {AIAA 2016-3518},
pubstate = {published},
tppubtype = {article}
}

@article{dorigo_toward_2022,
	title = {Toward the End-to-End Optimization of Particle Physics Instruments with Differentiable Programming: a White Paper},
	url = {http://arxiv.org/abs/2203.13818},
	shorttitle = {Toward the End-to-End Optimization of Particle Physics Instruments with Differentiable Programming},
	journal = {{arXiv}:2203.13818 [physics]},
	author = {Dorigo, Tommaso and others},
	longauthor = {Dorigo, Tommaso and Giammanco, Andrea and Vischia, Pietro and Aehle, Max and Bawaj, Mateusz and Boldyrev, Alexey and Manzano, Pablo de Castro and Derkach, Denis and Donini, Julien and Edelen, Auralee and Fanzago, Federica and Gauger, Nicolas R. and Glaser, Christian and Baydin, Atılım G. and Heinrich, Lukas and Keidel, Ralf and Kieseler, Jan and Krause, Claudius and Lagrange, Maxime and Lamparth, Max and Layer, Lukas and Maier, Gernot and Nardi, Federico and Pettersen, Helge E. S. and Ramos, Alberto and Ratnikov, Fedor and Röhrich, Dieter and de Austri, Roberto Ruiz and del Árbol, Pablo Martínez Ruiz and Savchenko, Oleg and Simpson, Nathan and Strong, Giles C. and Taliercio, Angela and Tosi, Mia and Ustyuzhanin, Andrey and Zaraket, Haitham},
	urldate = {2022-06-23},
	date = {2022-03-22},
	eprinttype = {arxiv},
	eprint = {2203.13818},
  year={2022},
}

@article{baydin_ad_in_ml_survey,
author = {Baydin, At\i{}l\i{}m G\"{u}nes and Pearlmutter, Barak A. and Radul, Alexey Andreyevich and Siskind, Jeffrey Mark},
title = {Automatic Differentiation in Machine Learning: A Survey},
year = {2017},
issue_date = {January 2017},
publisher = {JMLR.org},
volume = {18},
number = {1},
issn = {1532-4435},
journal = {J. Mach. Learn. Res.},
month = {jan},
pages = {5595–5637},
numpages = {43}
}

@article{aglieri_rinella_alpide_2017,
	title = {The {ALPIDE} pixel sensor chip for the upgrade of the {ALICE} {Inner} {Tracking} {System}},
	volume = {845},
	issn = {01689002},
	url = {https://linkinghub.elsevier.com/retrieve/pii/S0168900216303825},
	doi = {10.1016/j.nima.2016.05.016},
	language = {en},
	urldate = {2022-06-24},
	journal = {Nuclear Instruments and Methods in Physics Research Section A: Accelerators, Spectrometers, Detectors and Associated Equipment},
	author = {Aglieri Rinella, Gianluca},
	month = feb,
	year = {2017},
	pages = {583--587},
}

@manual{catphan-manual,
  title = {{Catphan\textsuperscript{\textregistered} 500 and 600 Manual}},
  key={{The Phantom Laboratory Inc., 2006}},
  organization={The Phantom Laboratory Inc.},
  address={PO Box 511, Salem, NY 12865-0511},
  year={2006},
}

@misc{aehle_forward-mode_2022,
  title = {Forward-{Mode} {Automatic} {Differentiation} of {Compiled} {Programs}},
  author = {Aehle, Max and Blühdorn, Johannes and Sagebaum, Max and Gauger, Nicolas R.},
  url = {http://arxiv.org/abs/2209.01895},
  publisher = {arXiv},
  month = sep,
  year = {2022},
  note = {arXiv:2209.01895 [cs]},
}

@misc{aehle_reverse-mode_2022,
  title = {Reverse-{Mode} {Automatic} {Differentiation} of {Compiled} {Programs}},
  author = {Aehle, Max and Blühdorn, Johannes and Sagebaum, Max and Gauger, Nicolas R.},
  url = {https://arxiv.org/pdf/2212.13760.pdf},
  publisher = {arXiv},
  month = dec,
  year = {2022},
  note = {arXiv:2212.13760 [cs]},
}

@techreport{urban,
      author        = "Urbán, László",
      collaboration = "GEANT4",
      title         = "{A model for multiple scattering in GEANT4}",
      institution   = "CERN",
      reportNumber  = "CERN-OPEN-2006-077",
      address       = "Geneva",
      year          = "2006",
      url           = "https://cds.cern.ch/record/1004190",
}

@article{FERNANDEZVAREA1993447,
title = {{On the theory and simulation of multiple elastic scattering of electrons}},
journal = {Nuclear Instruments and Methods in Physics Research Section B: Beam Interactions with Materials and Atoms},
volume = {73},
number = {4},
pages = {447-473},
year = {1993},
issn = {0168-583X},
doi = {https://doi.org/10.1016/0168-583X(93)95827-R},
url = {https://www.sciencedirect.com/science/article/pii/0168583X9395827R},
author = {J.M. Fernández-Varea and R. Mayol and J. Baró and F. Salvat},
}

@article{Ivanchenko_2010,
doi = {10.1088/1742-6596/219/3/032045},
url = {https://dx.doi.org/10.1088/1742-6596/219/3/032045},
year = {2010},
month = {apr},
publisher = {},
volume = {219},
number = {3},
pages = {032045},
author = {V N Ivanchenko and  O Kadri and  M Maire and  L Urban},
title = {{Geant4 models for simulation of multiple scattering}},
journal = {Journal of Physics: Conference Series},
}

@article{GIACOMETTI2017182,
title = {{Development of a high resolution voxelised head phantom for medical physics applications}},
journal = {Physica Medica},
volume = {33},
pages = {182-188},
year = {2017},
issn = {1120-1797},
doi = {https://doi.org/10.1016/j.ejmp.2017.01.007},
url = {https://www.sciencedirect.com/science/article/pii/S1120179717300078},
author = {V. Giacometti and S. Guatelli and M. Bazalova-Carter and A. B. Rosenfeld and R. W. Schulte},
}

\end{document}